\documentclass[twocolumn]{aastex6}
\shorttitle{Detecting Inclined Maser Disks Orbiting Massive Black Holes}
\shortauthors{Darling}

\begin{document}


\title{How to Detect Inclined Water Maser Disks (and Possibly Measure Black Hole Masses)}


\author{Jeremy Darling}
\affil{Center for Astrophysics and Space Astronomy,
Department of Astrophysical and Planetary Sciences,
University of Colorado, 389 UCB, Boulder, CO 80309-0389, USA}
\email{jdarling@colorado.edu}

\begin{abstract}
We describe a method to identify inclined water maser disks orbiting massive black holes
and to potentially use them to measure black hole masses.  
Due to the geometry of maser amplification pathways, the minority
of water maser disks are observable:  only those viewed nearly edge-on 
have been identified, suggesting that an order of magnitude additional maser disks exist.  
We suggest that inward-propagating masers will be gravitationally  
deflected by the central black hole, thereby scattering water maser emission 
out of the disk plane and enabling detection.  
The signature of an inclined water maser disk would be narrow masers near the systemic velocity that 
appear to emit from the black hole position, as identified by the radio continuum core.  
To explore this possibility, we present high resolution (0.07\arcsec--0.17\arcsec) 
Very Large Array line and continuum observations of 13 
galaxies with narrow water maser emission and show that three are good inclined disk candidates
(five remain ambiguous).  In the best case, for CGCG 120$-$039, we show that the maser and continuum 
emission are coincident to within $3.5\pm1.4$ pc ($6.7\pm2.7$ milliarcsec).
Subsequent very long baseline interferometric maps can confirm candidate inclined disks and
have the potential to show maser rings or arcs
that provide a direct measurement of black hole mass, although the mass precision will rely on knowledge
of the size of the maser disk.  
\end{abstract}




\section{Introduction} \label{sec:intro}

Water masers arising from thin disks around massive black holes provide 
high brightness temperature non-thermal dynamical tracers of gas in Keplerian 
orbits.  As such, water maser disks viewed edge-on provide tracers of the
Keplerian potential and enable measurement of the black hole mass, provided a distance 
is known in order to translate the apparent angular disk size into a physical 
size \citep[e.g.,][]{miyoshi1995}.
Maser accelerations and proper motions can also be observed, and because
the circular velocity is known from Doppler shifts, the geometric distance 
to the black holes (and host galaxies) can be determined \citep[e.g.,][]{herrnstein1998}.  
Geometric distances obtained from water masers provide a crucial independent measurement of the Hubble
constant and can be used to calibrate other distance indicators such as 
the period-luminosity relation of Cepheids \citep{riess2016}.

These measurements require maser disks that are viewed within a few degrees of edge-on:
otherwise, maser beaming directs emission away from the observer because masers 
propagate along velocity-coherent paths through the disk.  For thin disks, this propagation 
occurs along the radial path along the line of sight toward the black hole and along the disk 
tangent points.  For warped disks, such as that found in NGC 4258, the picture is
more nuanced because a warped disk provides numerous sightlines and inclinations that intersect velocity-coherent 
parts of the disk \citep{humphreys2013}.  Nonetheless, inclined maser disks are generally not seen in water maser
surveys.  Or are they?

\begin{figure*}[ht!]
\epsscale{1.1}
\plotone{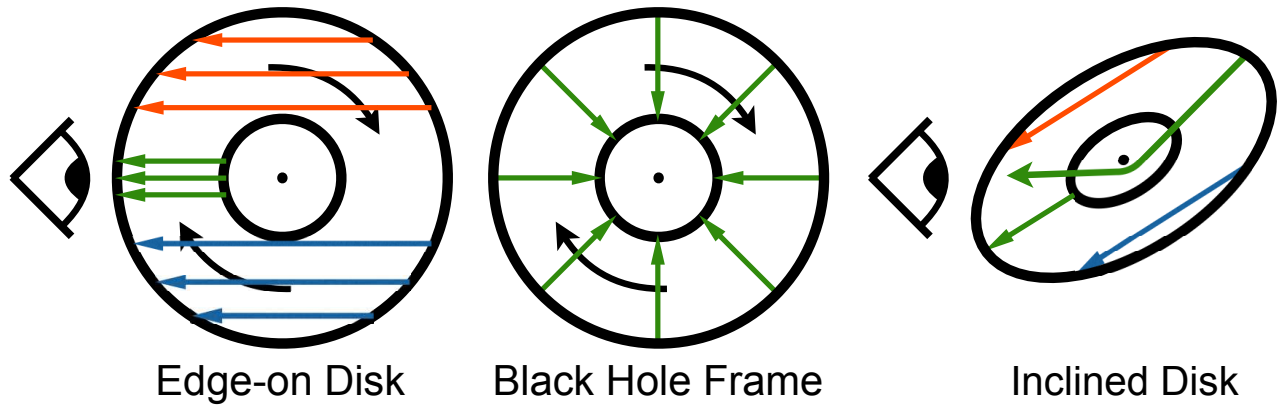}       
\caption{
Maser amplification pathways for Keplerian disks orbiting massive black holes.  These diagrams are schematic and not to scale.
Left:  the geometry of an egde-on rotating disk showing systemic velocity masers (green) and high-velocity redshifted (red) and blueshifted
(blue) masers at the tangent points of the disk.  
Center:  in the black hole rest frame, all disk motion is perpendicular to the radial direction, providing velocity-coherent amplification 
pathways at the systemic velocity (or rest frame).  In-going (and out-going) radial masers will populate the disk in this reference frame.
Right:  In-going masers may be gravitationally lensed/deflected by the central black hole into the sightline of an observer who would 
not otherwise see maser
emission from an inclined disk.   Only the systemic velocity masers would be seen by this observer, and they will appear to 
arise from the location of the black hole.  High-resolution imaging might reveal a ring or arc of water maser emission, 
providing a black hole mass measurement.  
\label{fig:schematic}}
\end{figure*}

It seems likely that inclined water maser disks have already been detected by single dish surveys, 
but they have been discarded because they show no high velocity lines that have canonically been
used to identify maser disks.  
In this paper, we propose a mechanism to produce detectable maser emission from inclined disks 
that may also be used to obtain black hole masses
(Section \ref{sec:thought}), we present a method to detect inclined maser disks based on extant surveys 
(Section \ref{sec:candidates}), we present Karl G. Jansky Very Large Array 
(VLA)\footnote{The National Radio Astronomy Observatory is a facility of the National Science Foundation operated under cooperative agreement by Associated Universities, Inc.} observations of candidate inclined disks 
(Sections \ref{sec:obs} and \ref{sec:results}), and we 
provide a list of inclined disk candidates for further study (Sections \ref{sec:analysis} and \ref{sec:discussion}).  
In what follows, we assume a flat cosmology with parameters  $H_0 = 70$~km~s$^{-1}$~Mpc$^{-1}$, $\Omega_M = 0.27$, and 
$\Omega_\Lambda = 0.73$, and we calculate distances from redshifts in the CMB rest frame.

\section{A Thought Experiment} \label{sec:thought}

Consider a typical 22 GHz water maser thin disk in Keplerian orbit around a massive black hole.
This thin disk shows systemic velocity masers at the center of the observed disk, along a radial amplification path, 
and it shows high-velocity red- and blue-shifted masers at the tangent points of the disk, along azimuthal paths, 
amplifying spontaneous emission and background host galaxy continuum (Figure \ref{fig:schematic}, left).  
If a thin disk is inclined by more than a few degrees, the masers are 
no longer beamed toward the observer and the disk would not be detected.  
The known water maser disks show an inclination typically within $\sim$5$^\circ$ of edge-on \citep[e.g.,][]{kuo2011}, 
suggesting that an order of magnitude additional maser disks exist, and they are simply beamed in directions we cannot observe.  
Note that the famous water maser disk in NGC 4258 \citep[e.g.,][]{miyoshi1995,herrnstein1998} is atypical, with 
a disk inclination of 72$^\circ$, but the disk warp provides sightlines that are nearly edge-on, and 
this is where the masers are seen \citep{humphreys2013}.

For the following discussion we will use fiducial parameters similar to those of many of the known maser disks 
\citep{kuo2011,greene2016}.  We assume a distance
of 10 Mpc, a black hole mass of $M_{BH} = 10^7\ M_\odot$ (and therefore a Schwarzschild radius of $R_s\sim1\ \mu$pc), and
a circular rotation speed of 1000 km s$^{-1}$ in the maser-active part of the disk located 0.5 pc from the black hole and 
spanning 0.2 pc.  We 
assume a disk thickness equal to that of NGC 4258: $5.5\times10^{14}$ cm or 0.18 mpc \citep[1$\sigma$;][]{argon2007}. 

Consider how this system appears to an observer at the location of the black hole:
the disk describes a plane in the sky.  Water masers reaching this observer must be purely radial (propagating inward) but will have a
coherent amplification path in all radial directions (Figure \ref{fig:schematic}, center).  All Keplerian Doppler shifts
will be across the line of sight, so all maser emission reaching this location will have zero velocity in this reference frame (ignoring
the transverse Doppler shift, 
which is negligible at $\sim$2 km~s$^{-1}$).  The black hole-frame observer will thus see a ring of maser light 
or maser spots with low velocity spread.  The maser-active part of the disk would subtend $\sim$2.5 arcminutes (1$\sigma$). 
The expected maser opening angle would be similar, $\sim$3 arcminutes, assuming a maser path-to-size ratio of 
$\sim$0.2 pc/0.2 mpc $\simeq 1000$, implying maser light spanning $\sim$500 $R_s$ at the black hole.   
 
The black hole Kerr metric would thus be bathed in maser light with a continuous distribution of impact parameters, 
implying gravitational deflection of maser light 
into nearly all angles (the deflection at 100 $R_s$ is 1.1$^\circ$ and grows inversely with impact parameter $R$ as $\theta = 2 R_s/R$;
\citet{einstein1915}).  
One would therefore
expect incoming maser light to be scattered into many angles, representing a gravitational lensing de-focusing of a given incoming
beam.  But portions of multiple maser beams from around the disk may be deflected into a given direction, making the 
net flux as a function of angular direction uncertain and in need of numerical modeling. 

NGC 4258 has a warped disk \citep[e.g.,][]{humphreys2013}, which means that the in-going radial masers can be misaligned and thus be
beamed above or below the dynamical center of the disk, and that the continuous distribution of disk inclination angles will 
fully populate the black hole metric volume with beams from multiple angles, 
further enhancing the sampling of impact parameters and deflection angles. 
If warped disks are common, 
then maser disks with high velocity lines will be detected more often than one would assume based on random inclination alone 
(in other words, parts of many 
disks will have inclinations appropriate for detectable maser amplification), and the black hole will be illuminated by in-going 
masers from above and below
the nominal disk plane, producing a range of incoming maser impact parameters and angles with respect to the disk,
thereby enhancing the gravitational deflection probability toward any given observer.  
Regardless of whether disks are warped or not, maser light will be 
scattered away from the maser disk plane, making inclined disks potentially detectable, but likely faint compared to edge-on disk
masers.

Masers amplify seed photons, and these seed photons can be the portion of a radio continuum that matches the maser line 
frequency, appropriately Doppler-shifted, or they can be spontaneously emitted maser line photons.  In known maser disks, the
systemic radial masers typically amplify the AGN radio continuum, and the high-velocity tangential masers amplify spontaneous 
emission or possibly radio continuum from the host galaxy itself.  In-going radial masers do not have access to AGN seed photons
and must amplify either continuum from the host galaxy or inward-directed spontaneous emission from the outer parts of the disk.  
In either case, it is reasonable to expect in-going masers to be weaker than the out-going systemic masers from 
edge-on disks.  It is unclear how the amplification of in-going masers would compare to the amplification of the 
high-velocity masers in edge-on disks because 
the tangential amplification pathway may or may not have a physically longer or higher column density 
velocity-coherent path for amplification.  

If one were to observe a water maser disk from a direction other than edge-on, would water maser emission be seen?  
The thought experiment 
described here suggests that it would, and the inclination of the disk with respect to the observer would select the deflection angle
(or equivalently the impact parameter) of the observed maser light.  For example, deflection by 10$^\circ$ would allow observation of masers
from the back side of a disk with an inclination of 80$^\circ$, requiring an impact parameter of 12 $R_s$.  Other parts of the disk could be viewed if light reaches smaller impact parameters.  If, as expected, there is a continuous range of impact parameters, then a continuous distribution of maser light from an extended portion of the disk will contribute to a spectrally narrow maser line complex seen by the observer at nearly any inclination (but perhaps with higher likelihood and intensity at higher inclinations).  

\subsection{Observable Signatures}  
 
The observational signatures of an inclined water maser disk would be:\\
\indent 1.  A narrow line or line complex\\
\indent 2.  at the systemic velocity\\
\indent 3.  at the apparent black hole location.\\
The black hole location would be indicated by the radio continuum core, ideally observed at the same frequency as the water
maser.
The observational signatures of an inclined water maser disk, however, may also arise from other mechanisms.
Water masers can be produced in radio jet-molecular cloud interactions \citep[e.g.,][]{gallimore1996,claussen1998,peck2003,henkel2005}, in 
star-forming regions 
\citep[e.g.,][]{tarchi2002a,tarchi2002b,henkel2005,hofner2006, darling2008,brogan2010,darling2011,tarchi2011,amiri2016}, 
and in outflows \citep[e.g.,][but note that some objects in the latter survey may be 
candidates for inclined disk masers]{greenhill2003b,kondratko2005,tarchi2011a}.
These are likely to be the main contaminant among an inclined disk
survey sample.  VLBI identification of the maser with an AGN via spatial coincidence with 
the core radio continuum --- identified by the spectral index --- can resolve the ambiguity (Section \ref{sec:results}).

In contrast to edge-on disk systemic masers, it is unclear whether inclined disk maser lines will show proper motion or acceleration.  This may depend on the clumpiness of maser-emitting regions, on the maser beam sizes, on the physical extent of the 
disk that is sampled by the observed line, and on the deflection angle to the observer.
One might expect lensed masers to show little or no time variability, but the substructure seen in known water maser disks 
and the natural variability of water masers suggests that this may be a bad assumption.  

Many extragalactic water masers detected in previous surveys meet some or all of the above observational criteria:  
inclined water maser disks may have already been detected!  In most cases, when 
a single systemic velocity line is detected in water maser surveys, there is no interferometric follow-up
because distance or black hole mass measurements (traditionally) require edge-on disks, the signature of which are 
the high velocity lines emitted from the tangent points of the disk.  
High resolution observations may also be frustrated by weak or variable water masers.

In Section \ref{sec:obs} we present interferometric mapping of a sample of narrow-line systemic velocity water masers that appear to be inclined disk maser candidates.  If observations show that the masers remain unresolved and are centered at the location of the 
central massive black hole (as identified by simultaneous radio continuum observations at 20 GHz), then they remain candidates and should be mapped with VLBI.

\subsection{Black Hole Masses}

The Einstein radius of a strong gravitational lens is 
\begin{equation}
   \theta_E = \sqrt{2 R_s {D_{LS}\over D_L D_S}} 
                 = 9.2 \sqrt{{R_s \over {\rm mpc}} \, {D_{LS} \over {\rm pc}}} \left( D_L \over {\rm Mpc} \right)^{-1} {\rm mas}, 
\end{equation}
where $D_L$, $D_S$, and $D_{LS}$ are the angular diameter distances to the lens, to the source, and between the lens and the source,
respectively, and $D_L \simeq D_S$ for the black hole-maser disk configuration \citep[after][]{einstein1936}.
For the fiducial parameters listed above ($R_s = 1$ $\mu$pc, $D_{LS} = 0.5$ pc, and $D_L = 10$ Mpc),
$2 \theta_E = 0.041$ mas. 
Since the angular resolution (HPBW) of the Very Long Baseline Array (VLBA) is 0.3 milliarcseconds at 22.2 GHz (which should be 
compared to twice the Einstein radius), 
the Einstein radius for the fiducial maser disk and black hole would require space-based VLBI to resolve.  
On the other hand, terrestrial VLBI could resolve 
a water maser Einstein ring for a more massive black hole with a physically larger maser disk:  for 
$R_s = 100$ $\mu$pc ($M_{BH} = 10^9\ M_\odot$), $D_{LS} = 2$ pc, and $D_L = 10$ Mpc,
$2 \theta_E = 0.82$ mas.  Unfortunately, water maser disks have yet to be identified orbiting $10^9\ M_\odot$ black holes, 
but it is unclear whether is this a selection effect or a consequence of physics \citep{vandenbosch2016}.  The most 
massive black hole measured using a water maser disk to date is in NGC 1194 with $M_{BH} = 10^{7.85\pm0.05}\ M_\odot$ \citep{kuo2011,greene2016}.

Were an Einstein ring observable from a back side in-going maser in an edge-on disk or from an in-going maser from 
an edge-on portion of an inclined warped disk, then one can infer a black hole mass from the angular size of the ring.  
One does need independent measurements of the luminosity distance to the black hole ($D_L$) and the size of the 
maser-emitting disk ($D_{LS}$).  $D_L$ can be obtained from an assumed cosmology and the cosmological redshift, but 
$D_{LS}$ may be more difficult to measure.  The precision of black hole masses obtained from this method may be limited
by our ability to measure, model, or estimate the radius of the maser-emitting part of the disk.  

Einstein rings require linear alignment between the source, the lens, and the observer, which is not germane to the inclined
maser disk geometry (for unwarped disks). 
Instead, one would na\"{i}vely expect to see multiple images of the same maser, which can also be related to the black hole mass.  This
expectation, which is correct for isotropic emitters, is probably incorrect for masers.

An important difference between maser emission and the standard treatment of gravitational lensing is maser 
beaming:  while there may be sightlines in a gravitational lens geometry that land on the emitter, emission 
may not be seen if light is not beamed along that sightline.  For a general pointlike isotropic light source offset from 
the lens by angle $\theta_S$, there are two solutions to the lens equation:
\begin{equation}
  \theta_\pm = {1\over2} \left( \theta_S \pm \sqrt{\theta_S^2 + 4 \theta_E^2}\right).
\end{equation}
$\theta_+$ represents the angle between the lens and the source image appearing outside $\theta_E$, and 
$\theta_-$ represents the angle of the image appearing inside $\theta_E$
\citep[][Equation 24, Figures 5 and 7]{narayan1995}.
For an inclined maser disk configuration, no emission is directed 
significantly out of the disk plane, so no maser emission would be seen in the $\theta_+$ direction.
On the other hand, $\theta_-$ may be small enough that this sightline is included in the maser beam 
passing very close to the central black hole.  In this case, the source-observer deflection angle $\alpha_-$ nearly matches the 
complement of the disk inclination:  $\alpha_- \simeq \pi/2 - i$ (Figure \ref{fig:schematic}, right).  

In contrast to canonical lensing, 
we expect that the maser beaming will produce only one maser spot image, where light is deflected in the 
manner shown schematically in Figure \ref{fig:schematic} (right).  This maser image will lie inside the Einstein radius.
For isotropic extended emitters, this image would also be demagnified, but masers are not isotropic emitters and 
can be exceptionally compact (equivalently, they demonstrate high brightness temperatures).  The degree of demagnification
is therefore unclear and requires numerical ray-tracing to assess.

For the fiducial parameters above, and assuming an inclination of $80^\circ$,
$\theta_S = (R_{\rm maser}/D_S) \sin\alpha_- = 1.8$ mas, $\theta_E = 0.021$ mas, and therefore $|\theta_-| = 0.2$ $\mu$as, 
which is equivalent to 11 $R_s$.  The maser beam spans this distance from 
the black hole, so the maser emission can be lensed toward the observer in this case. 
For the $10^9\ M_\odot$ black hole with a larger maser disk described above, 
$\theta_S = (R_{\rm maser}/D_S) \sin\alpha_- = 7.2$ mas,
$\theta_E = 0.41$ mas, and therefore $|\theta_-| = 24$ $\mu$as, 
which is again equivalent to 11 $R_s$ (the impact parameter determines the 
deflection angle, which is determined by the inclination).  

One would therefore expect lensed masers from inclined disks to be faint and appear to arise from 
the black hole location ($\theta_- \ll 1$ mas).   
This treatment assumes a single pointlike maser rather than an extended continuous disk of masers or a set of distributed maser 
spots.  In this more realistic scenario, numerical ray-tracing is required to connect the observable lensed maser image to the black 
hole mass and maser disk configuration.  

Lensed masers from inclined disks may appear to be pointlike or they may describe arcs.
In-going masers from the far side of an edge-on disk or in-going masers from an inclined but warped disk may produce Einstein
rings.  
Single-epoch VLBI maps can therefore provide black hole masses, but space-based VLBI may be required
for the typical $\sim10^7\ M_\odot$ black hole associated with water maser disks.   The mass measurement precision will be limited by 
uncertainty about the maser disk size rather than by the distance to the object
 (the maser will provide the systemic redshift, which can be converted into a distance, and
the distance uncertainty will be dominated by peculiar velocity departures from the Hubble flow).

\subsection{Back-Side Masers in Edge-on Disks}  

If there are in-going radial water masers in maser disks (and there is no compelling reason to think otherwise), 
then there should be systemic water masers seen in edge-on disks from the back side of the disk, 
both lensed and unlensed by the black hole.  
Are such masers in extant data?  

If there are also front-side systemic masers in an edge-on disk, then they will be orders of magnitude brighter 
than back-side masers, even in the presence of strong lensing, simply due to amplification considerations.  
The front-side masers can amplify AGN radio continuum, whereas the back-side masers will either be driven by 
stimulated emission (but with an amplification pathway equal to the front-side masers) or by host galaxy continuum, 
which is substantially weaker than AGN continuum at 22 GHz in many cases.   The back-side maser contribution may therefore
be confused by the front-side emission, particularly since both types of maser emission will occur at the systemic velocity.

In rare cases, a back-side maser might be distinguishable from the front-side emission either by a position or a velocity offset
(there is some spread to systemic velocity masers, e.g. \citet{gao2016}).  The observational signature of a back-side maser would be an acceleration 
in the opposite sense of the front-side systemic maser acceleration (i.e., negative acceleration under the convention that 
positive velocities are redshifted).   This is not seen in published systemic maser acceleration measurements 
\citep[e.g.][]{greenhill1995,nakai1995,braatz2010,kuo2015,gao2016}, but such a signal could be lost amid the brighter and numerous front-side masers.

If a back-side maser is lensed into an arc or Einstein ring by the central black hole, then it might be extended in VLBI maps.  Resolved 
systemic maser emission would therefore be an additional observable signature of back-side masers.

\section{Candidate Selection}\label{sec:candidates}

If inclined water maser disks can be detected via gravitational lensing or deflection of in-going masers by massive black holes, then 
they have likely already been detected in surveys for maser disks.  But they were rejected as disk candidates because they 
lacked high-velocity emission.  Inclined maser disks will appear to have maser
emission only at the systemic velocity of the galaxy or AGN.  We therefore use extant water maser surveys to select inclined disk 
candidates.

Using extant water maser surveys, most of which favor Seyfert 2 AGN, we examined single-dish spectra 
compiled by the Megamaser Cosmology Project\footnote{\label{footnote:maser_url}\url{https://safe.nrao.edu/wiki/bin/view/Main/PrivateWaterMaserList}}
to select objects showing
a narrow systemic velocity maser or maser complex.  We also imposed a 30 mJy line flux limit and excluded objects south of $-20^\circ$ declination.  
This process identified 16 inclined maser disk candidates (Table \ref{tab:obs}), and most candidates (14) have only been observed
with a single dish.  Those that do have interferometric maps are NGC 3556, which was mapped using the VLA in CnB and DnA configurations \citep{tarchi2011} with no 22 GHz continuum detected (1$\sigma$ rms noise of $\sim$0.5 mJy beam$^{-1}$),  
and NGC 3735,which was mapped using A-array \citep{greenhill1997}, but no 22 GHz continuum was detected.

Since the inclinations of known maser disks are nearly edge-on, the number of inclined maser disks must be large, 
roughly an order of magnitude larger than the number of detected maser disks.
However, the gravitational lensing or deflection of detectable in-going radial
masers adds substantial uncertainty to the detection expectations.   We do not know how common observable lensed inclined maser
disks are in the universe because we do not know the opening angle of the masers, the 
size of the maser spots with respect to the black hole's Schwarzschild radius, the brightness of in-going masers, which will 
depend on the 22 GHz seed photons from the host galaxy, or the degree of gravitational lensing demagnification.

\section{Observations and Data Reduction}\label{sec:obs}

We observed the 22.23508 GHz $6_{16}-5_{23}$ ortho water maser line and 20 GHz radio continuum 
toward 16 candidate inclined maser disks 
(see Section \ref{sec:candidates} and Table \ref{tab:obs}) using the 
VLA in A configuration (the highest angular resolution configuration).  Observations 
of program 15A-297 spanned June 19 2015 through September 26 2015 in five sessions.  The 
fifth session occurred during the A-array to D-array reconfiguration.   Each session included visits to 
flux and bandpass calibrators, and observations of each target object were interleaved with nearby 
complex gain calibrators with a $\sim$4 minute switching cadence.  

Spectral line and continuum observations were simultaneous.  The spectral line observations were centered on the redshift of the
host galaxy, had 1.1--1.9 km s$^{-1}$ spectral resolution, used 1536 channels to span 128 MHz (1700--2900 km s$^{-1}$), 
and used dual circular polarization
and 8-bit sampling.  Continuum observations spanned 4 GHz, 18--22 GHz, using 32 spectral windows spanning 128 MHz each 
using 128 channels in dual circular polarization and 3-bit sampling.  

Table \ref{tab:obs} lists the details of the observations and the rms noise in the line cubes and continuum maps.  Typical beam sizes were
80--100 milliarcseconds. Noise was about 3 mJy beam$^{-1}$ in 1.2 km~s$^{-1}$ channels in the spectral line cubes and 
about 15--20 $\mu$Jy beam$^{-1}$ in the continuum.  The exception was the $z\simeq0.66$ water maser J0804+3607 
\citep{barvainis2005}
that was redshifted to 13.4 GHz, in Ku band, which necessarily had lower angular and spectral resolution and lower rms noise
in the line but higher rms continuum noise.  In this case, the continuum was centered on 14 GHz and spanned 12--16 GHz.

The observing session on September 15 2015 had poor 22 GHz weather for A configuration and could not be calibrated or imaged.  
NGC 3359, NGC 3556, and NGC 3735
are therefore only listed in Table \ref{tab:obs} and are not discussed further or included in any analysis of the remaining 13 objects.

All data reduction and analysis was performed using the Common Astronomy Software Applications package \citep[CASA;][]{mcmullin2007}.  Calibration and flagging 
used a modified CASA pipeline plus additional manual flagging.  Imaging used Briggs weighting with robustness 0.5.  
Spectra were extracted from spectral line cubes using a maser-centered beam, and integrated line maps were restricted to 
line-emitting channels.  All spectra use the optical velocity definition in the Barycentric reference frame.

\floattable
\begin{deluxetable}{cccCrCccc}
\tablecaption{Journal of Observations \label{tab:obs}}
\tablecolumns{9}
\tablewidth{0pt}
\tablehead{
\colhead{Galaxy} & 
\colhead{UT Date} &
\colhead{Integration} & 
\multicolumn{3}{c}{Beam\tablenotemark{a}} & 
\multicolumn{2}{c}{Line} &
\colhead{Continuum} \\
\cline{4-6}
\colhead{} & 
\colhead{} & 
\colhead{} & 
\colhead{Angular} & 
\colhead{PA} & 
\colhead{Physical} & 
\colhead{rms} & 
\colhead{$\Delta v$} & 
\colhead{rms} \\
\colhead{} & 
\colhead{} & 
\colhead{(m)} & 
\colhead{(milliarcsec)} & 
\colhead{($^\circ$)} & 
\colhead{(pc)} & 
\colhead{(mJy bm$^{-1}$)} &
\colhead{(km s$^{-1}$)} &
\colhead{($\mu$Jy bm$^{-1}$)}  
}
\startdata
NGC 291          & 2015-06-21 & 23.3  &  128\times84 & 20.4 & 50\times33 & 2.6 & 1.2 & 17  \\ 
NGC 520b        & 2015-06-21 & 23.3  &  102\times89 & 27.8 & 14\times12 & 3.1 & 1.1 & 20 \\ 
J0350$-$0127 & 2015-06-21 & 23.3  &  100\times87 & 20.5 & 81\times70 & 3.1 & 1.2 & 15  \\ 
IC 485              & 2015-09-26 & 23.4  &   84\times82 & 19.6 & 47\times46 & 2.2 & 1.2 & 18  \\ 
J0804+3607 & 2015-09-26 & 23.3 & 151\times130 & $-$81.3 & 1065\times917 & 0.9 & 1.9 & 47   \\ 
CGCG 120$-$039 & 2015-09-26 & 23.3  & 85\times81 & 33.7 & 44\times42 & 2.1 & 1.2 & 20  \\ 
J0912+2304    & 2015-09-06 & 23.3  & 92\times81 & 65.6 & 67\times59 & 2.9 & 1.2 & 16  \\ 
J1011$-$1926 & 2015-09-06 & 23.4  & 173\times86 & $-$20.8 & 98\times49 & 5.3 & 1.2 & 19  \\ 
NGC 3359\tablenotemark{b}       & 2015-09-15 & 23.3  & 112\times72 & 48.6 & 9\times6 & \nodata & 1.1 & \nodata \\ 
NGC 3556\tablenotemark{b}       & 2015-09-15 & 23.3  & 104\times72 & 45.0 & 6\times4 & \nodata & 1.1 & \nodata  \\ 
NGC 3735\tablenotemark{b}       & 2015-09-15 & 23.3  & 112\times68 & 36.8 & 21\times13 & \nodata & 1.2 & \nodata  \\ 
UGC 7016       & 2015-09-06 & 23.3  & 114\times91 & $-$74.8 & 54\times43 & 3.8 & 1.2 & 16  \\ 
NGC 5256       & 2015-06-19 & 23.3  & 96\times84 & $-$73.8 & 55\times48 & 3.0 & 1.2 & 17 \\ 
NGC 5691       & 2015-06-19 & 23.4  & 131\times85 & 39.3 & 15\times9 & 3.7 & 1.1 & 17 \\ 
CGCG 168$-$018 & 2015-06-19 & 23.3 & 86\times82 & 50.0 & 63\times60 & 2.7 & 1.2 & 15  \\ 
J1939$-$0124 & 2015-06-19 & 25.0 & 96\times87 & 37.7 & 41\times38 & 3.0 & 1.2 & 16 \\ 
\enddata
\tablenotetext{a}{Synthesized beam properties for the spectral line observations.  The continuum maps are slightly different.}
\tablenotetext{b}{Objects observed on 2015 September 15 could not be calibrated or imaged due to poor weather.  }
\end{deluxetable}

\section{Results}\label{sec:results}

\begin{figure*}
\includegraphics[width=0.535\textwidth]{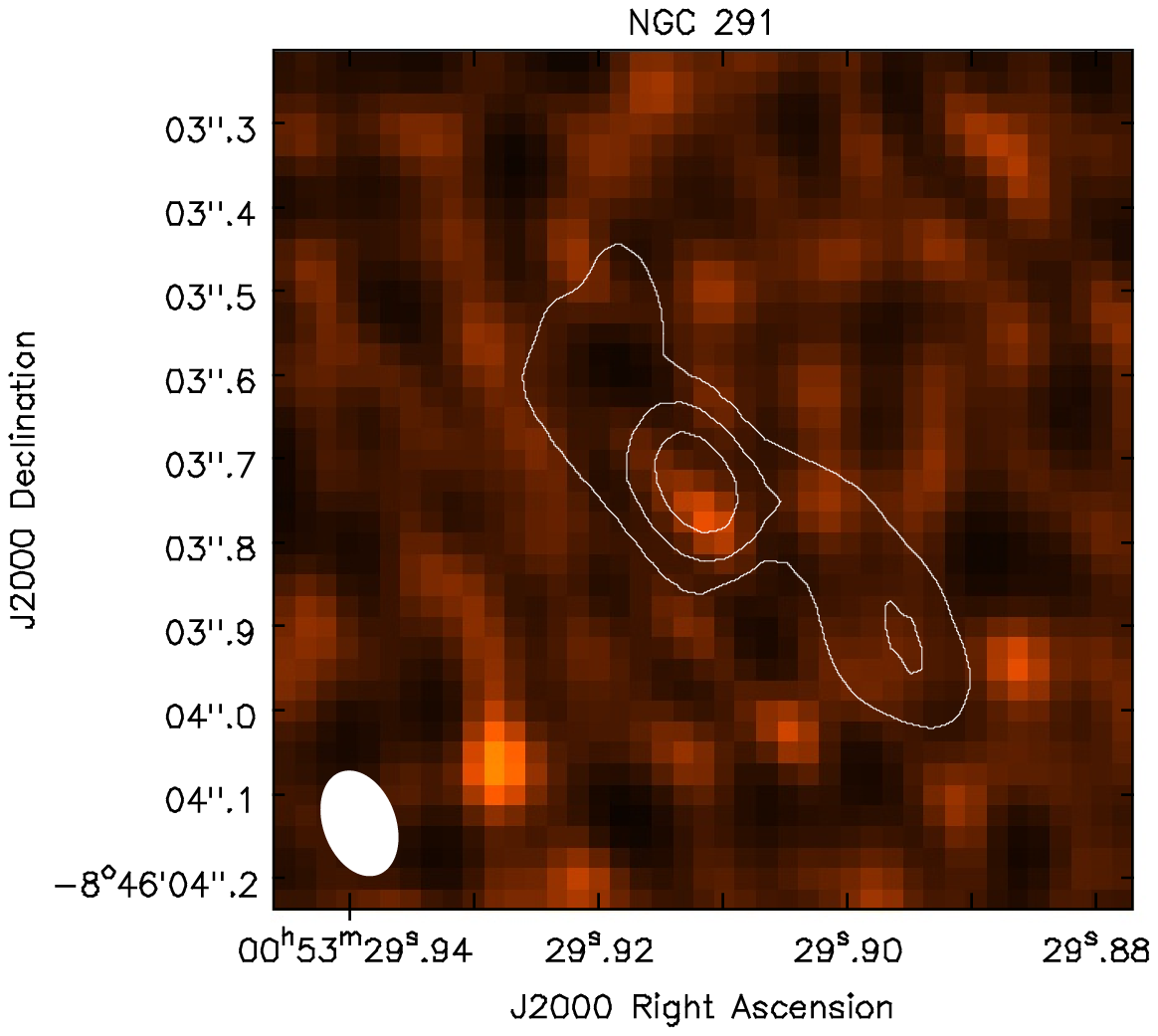}
\includegraphics[width=0.46\textwidth]{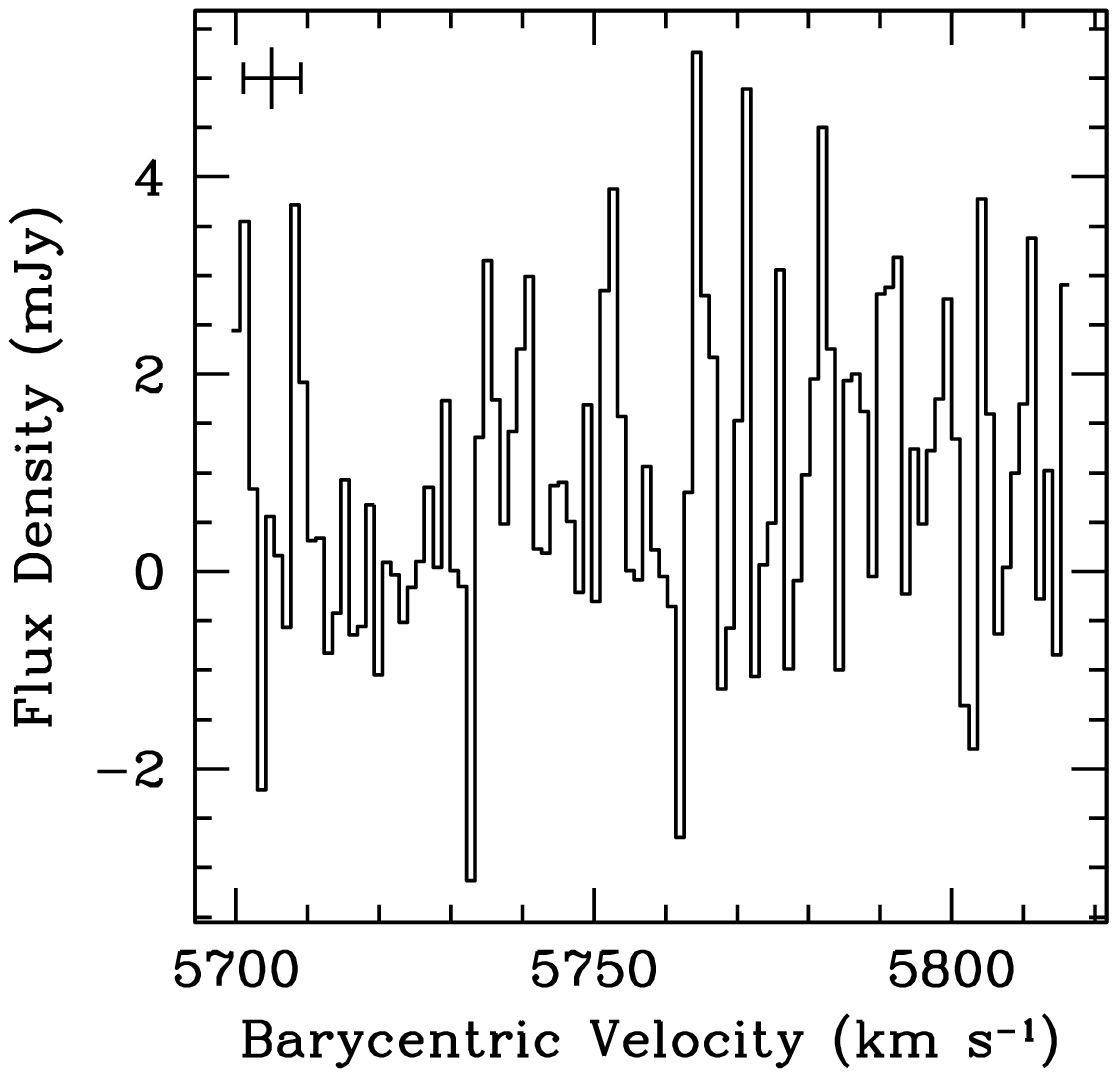}
\caption{Left:  NGC 291 integrated water maser (image) and 20 GHz radio continuum  (contours) maps.  Continuum contours 
indicate 4, 8, and 16 times the rms noise listed in Table \ref{tab:obs}.   The spectral line beam is shown in the lower left
(properties listed in Table \ref{tab:obs}). The 1\arcsec\ field of view is equivalent to 393 pc.  
Right:  water maser nondetection spectrum at the continuum peak with the systemic velocity and its uncertainty indicated by the vertical bars
\citep[$5705\pm4$ km s$^{-1}$;][]{SDSSDR2}.
The spectrum is roughly centered on the previous single-dish water maser detection (see Section \ref{sec:discussion}).  
\label{fig:NGC291}}
\end{figure*}

\begin{figure*}
\includegraphics[width=0.53\textwidth]{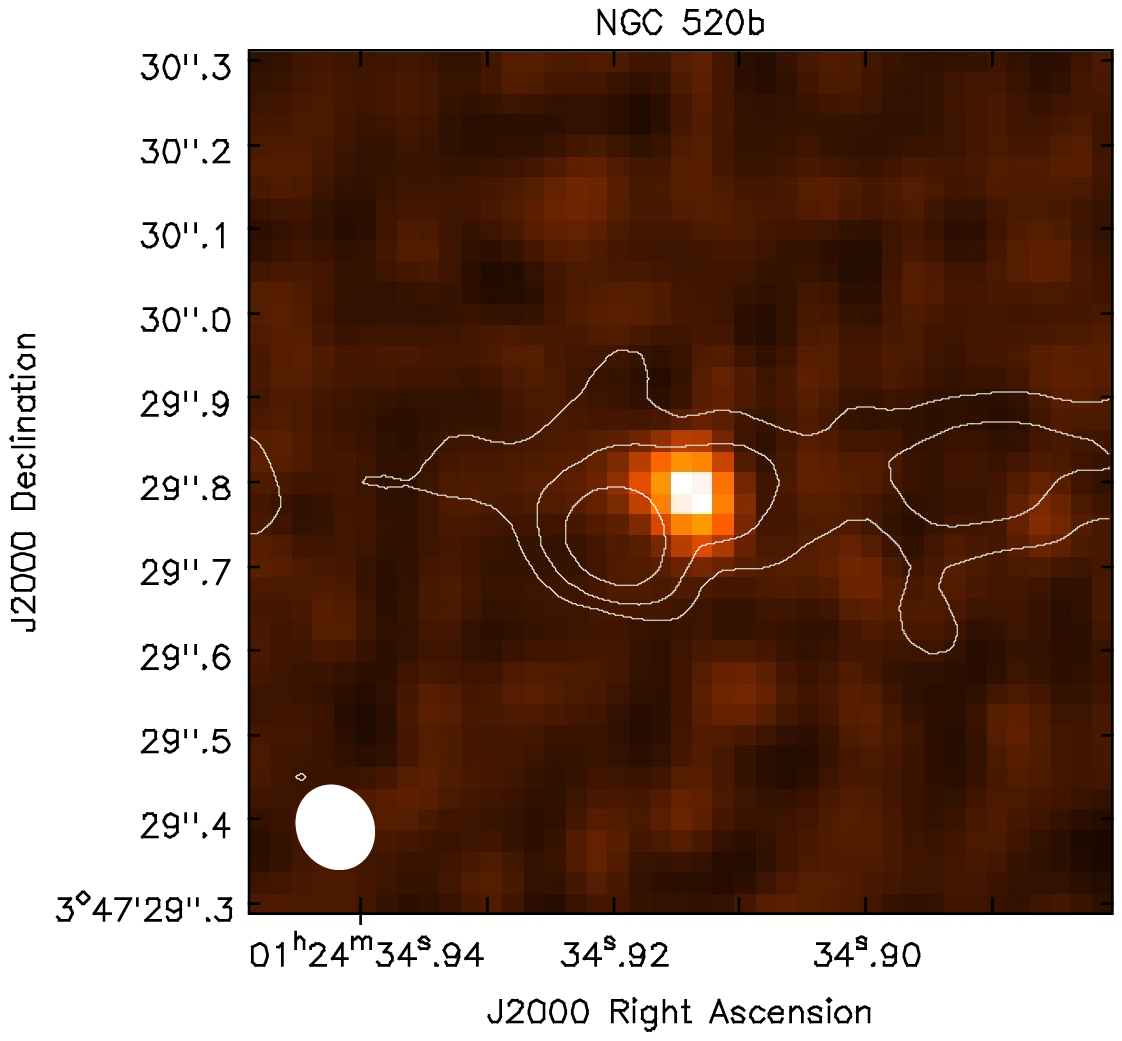}
\includegraphics[width=0.465\textwidth]{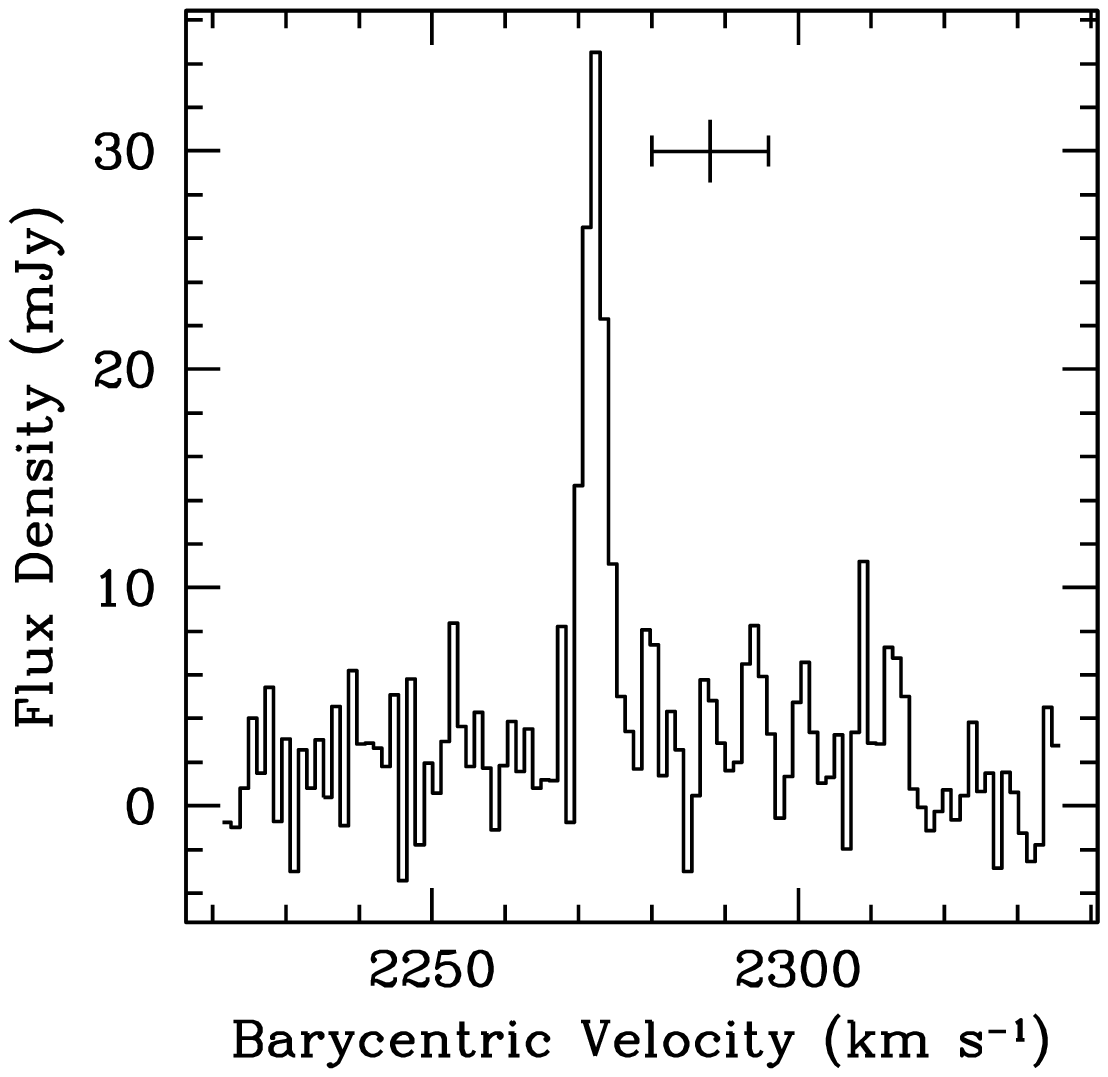}
\includegraphics[width=1.0\textwidth,trim={0 0 0 0},clip]{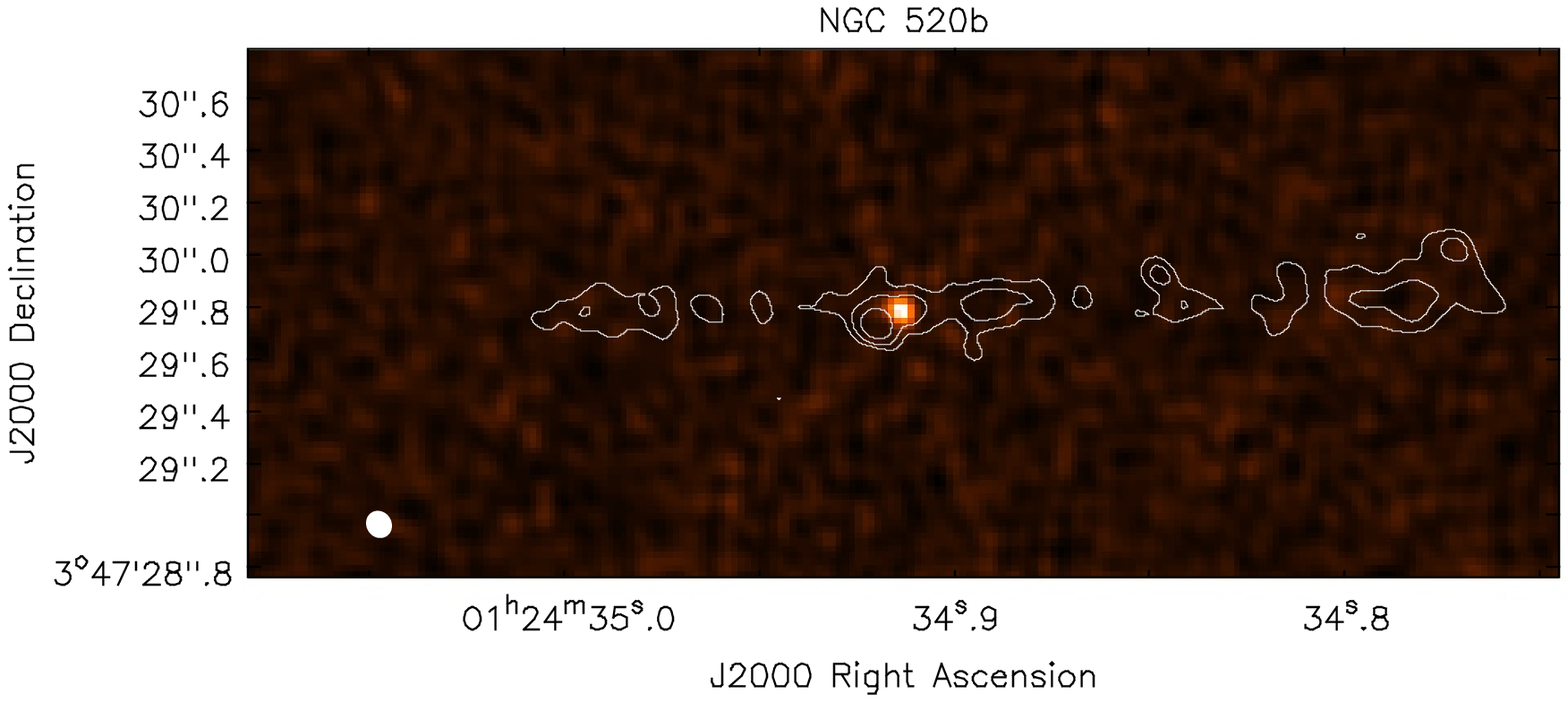}
\caption{Top Left:  NGC 520b integrated water maser (image) and 20 GHz radio continuum (contours) maps.  Continuum contours 
indicate 4, 8, and 16 times the rms noise listed in Table \ref{tab:obs}.   The spectral line beam is shown in the lower left
(properties listed in Table \ref{tab:obs}). The 1\arcsec\ field of view is equivalent to 135 pc.  
Top Right:  water maser spectrum with the systemic velocity and its uncertainty indicated by the vertical bars  (Table \ref{tab:masers}).  
The spectrum is roughly centered
on the previous single-dish water maser detection (see Section \ref{sec:discussion}).  
Bottom: a 5\arcsec$\times$2\arcsec\ (675 pc $\times$ 170 pc) field of view to show the full extent of the 20 GHz continuum.
\label{fig:NGC520b}}
\end{figure*}

\begin{figure*}
\includegraphics[width=0.52\textwidth]{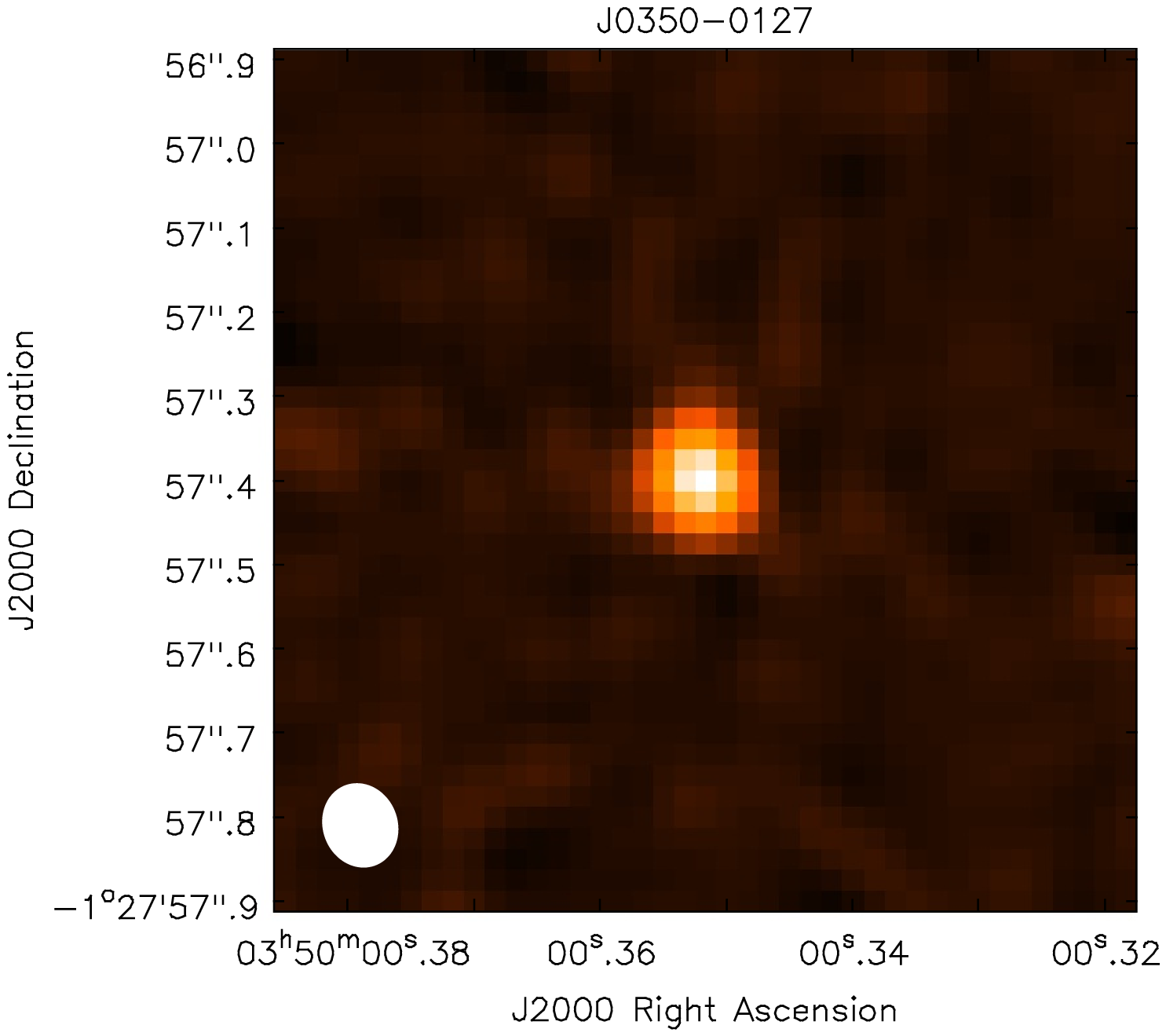}
\includegraphics[width=0.475\textwidth]{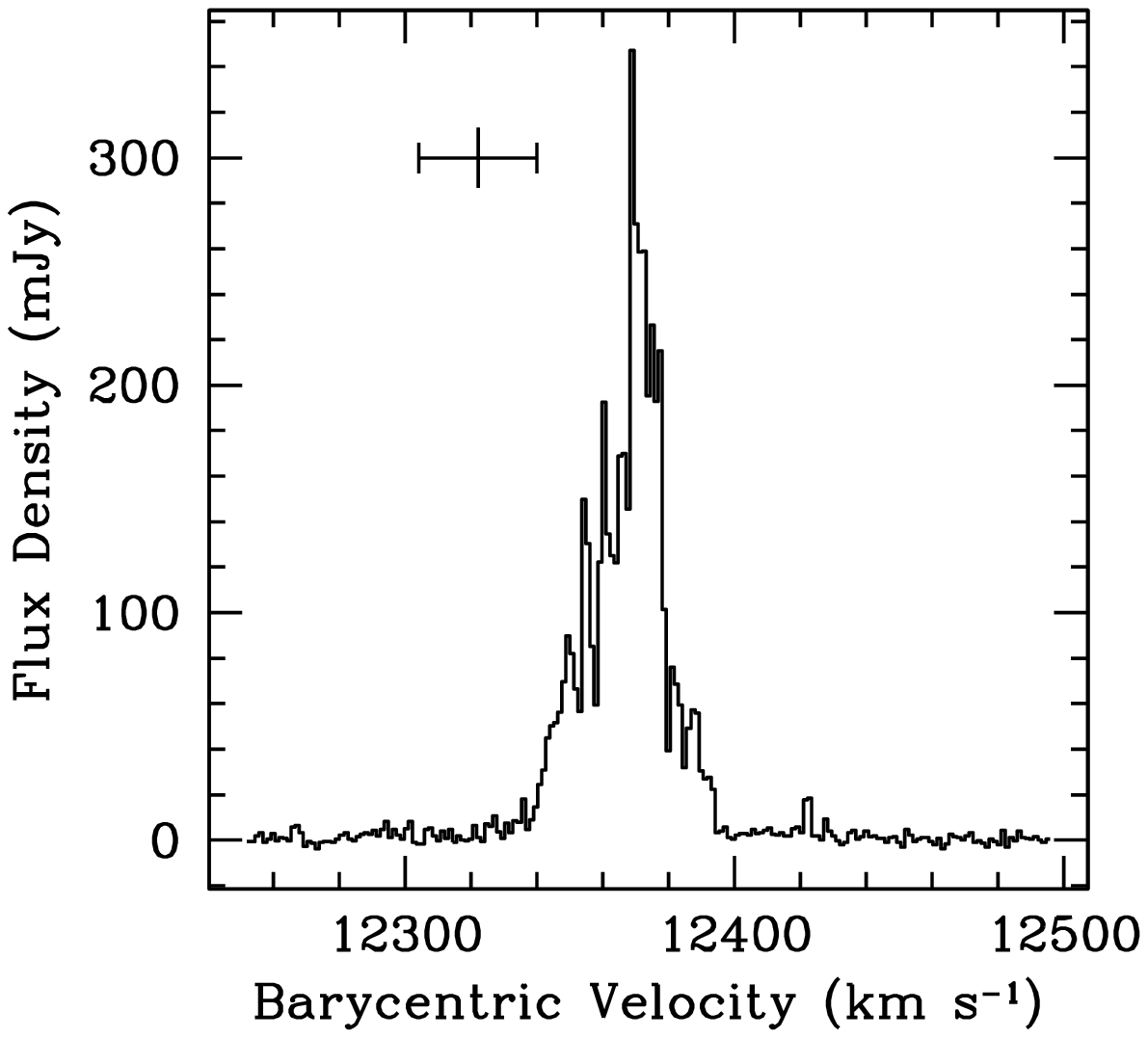}
\caption{Left:  J0350$-$0127  integrated water maser map.  The 20 GHz continuum was not significantly detected.
The spectral line beam is shown in the lower left
(properties listed in Table \ref{tab:obs}). The 1\arcsec\ field of view is equivalent to 805 pc.  
Right:  water maser spectrum with the systemic velocity and its uncertainty indicated by the vertical bars (Table \ref{tab:masers}).  
The spectrum is roughly centered on the previous single-dish water maser detection (see Section \ref{sec:discussion}).  
\label{fig:2MASX0350}}
\end{figure*}

\begin{figure*}
\includegraphics[width=0.525\textwidth]{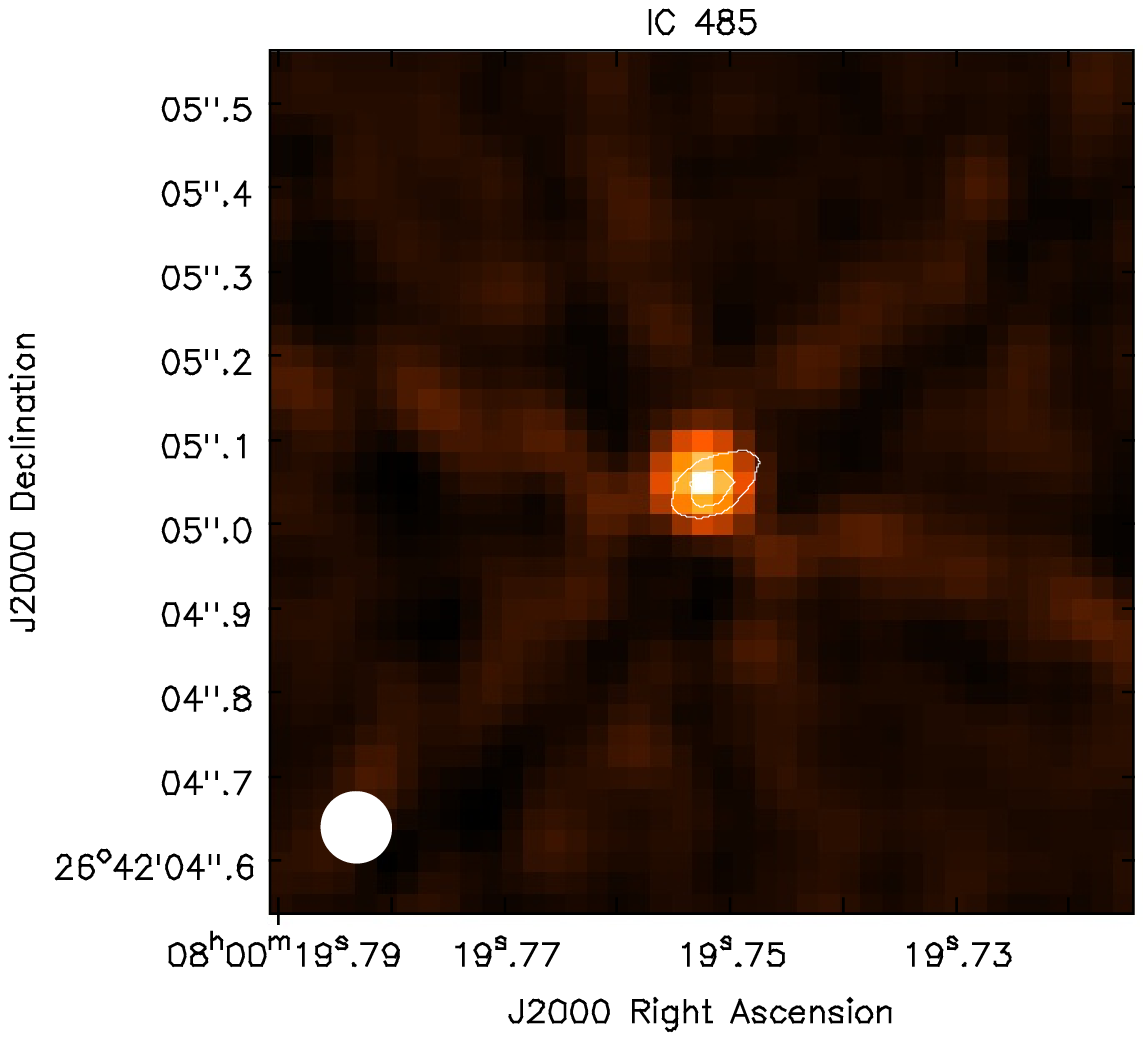}
\includegraphics[width=0.47\textwidth]{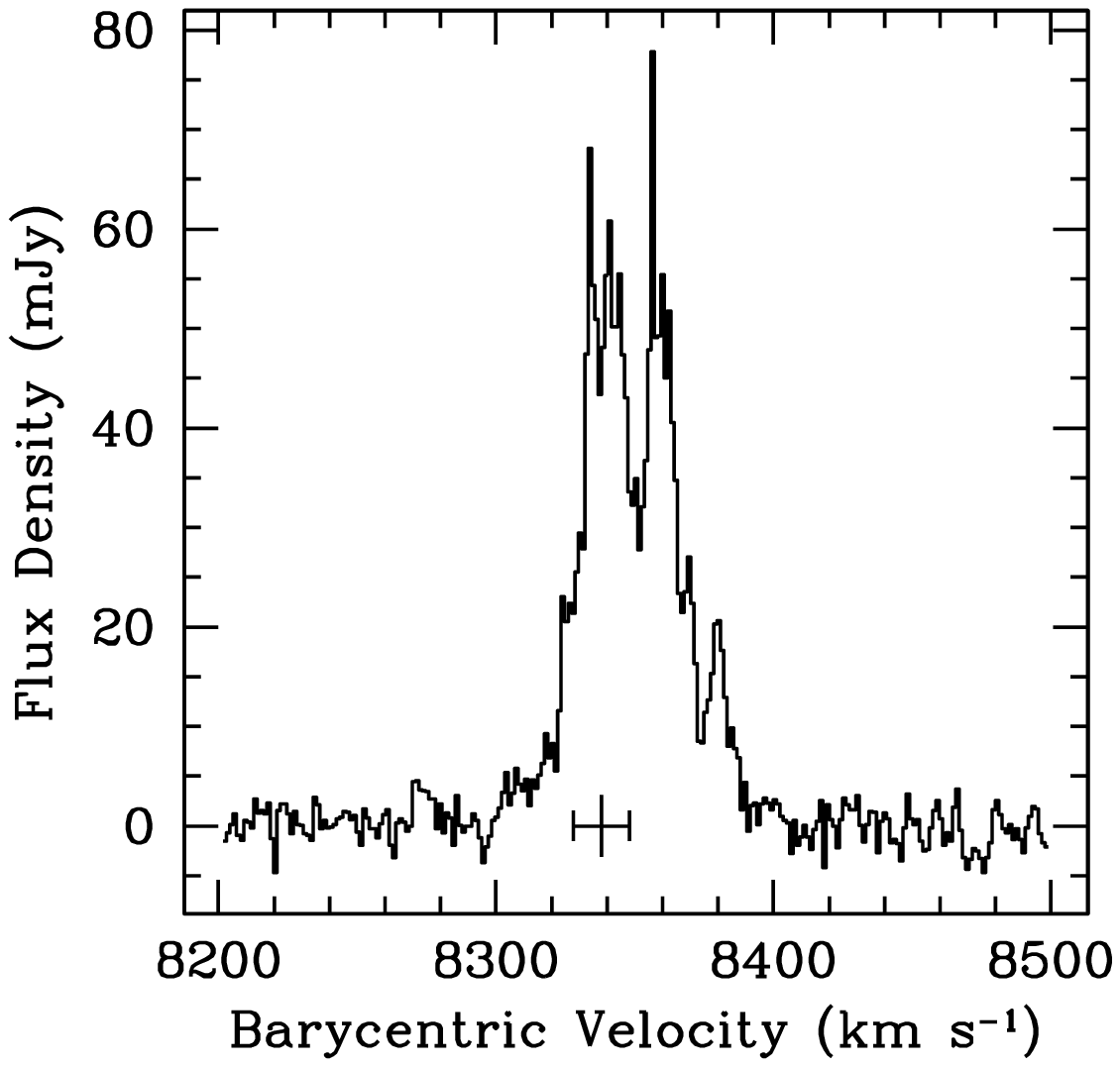}
\caption{Left: IC 485 integrated water maser (image) and 20 GHz radio continuum  (contours) maps.  Continuum contours 
indicate 3 and 4 times the rms noise listed in Table \ref{tab:obs}.   The spectral line beam is shown in the lower left
(properties listed in Table \ref{tab:obs}). The 1\arcsec\ field of view is equivalent to 562 pc.  
Right:  water maser spectrum with the systemic velocity and its uncertainty indicated by the vertical bars (Table \ref{tab:masers}). 
The spectrum is centered on the previous single-dish water maser detection (see Section \ref{sec:discussion}).  
\label{fig:IC485}}
\end{figure*}

\begin{figure*}
\includegraphics[width=0.53\textwidth]{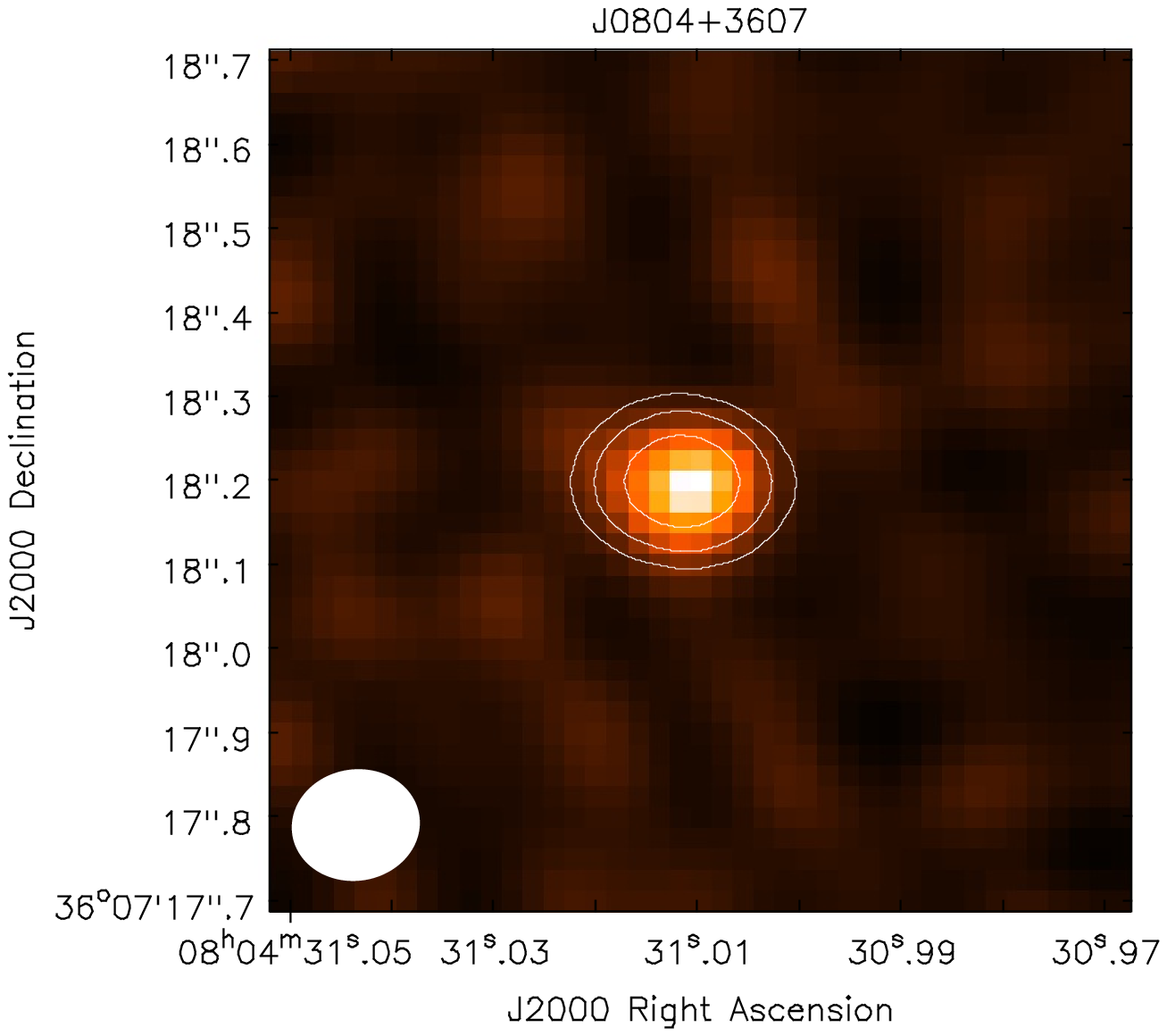}
\includegraphics[width=0.465\textwidth]{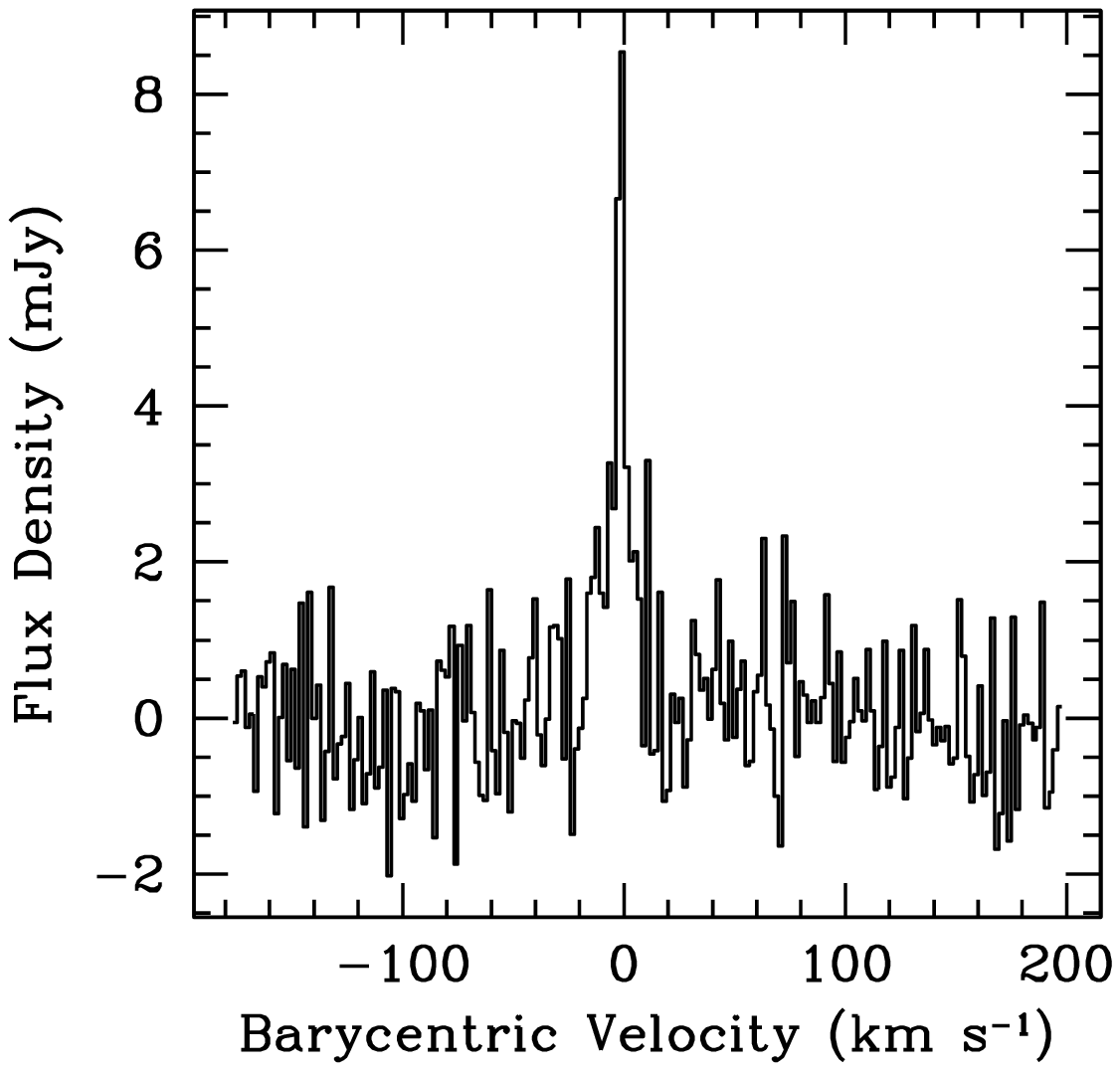}
\caption{Left:  J0804+3607 integrated water maser (image) and 14 GHz radio continuum  (contours) maps.  Continuum contours 
indicate 4, 8, and 16 times the rms noise listed in Table \ref{tab:obs}.   The spectral line beam is shown in the lower left
(properties listed in Table \ref{tab:obs}). The 1\arcsec\ field of view is equivalent to 7.055 kpc.  
Right:  water maser spectrum centered on the peak maser emission at $z=0.66045$, plotted in the object rest frame (Table \ref{tab:masers}).  
\label{fig:SDSSJ0804}}
\end{figure*}

\begin{figure*}
\includegraphics[width=0.53\textwidth]{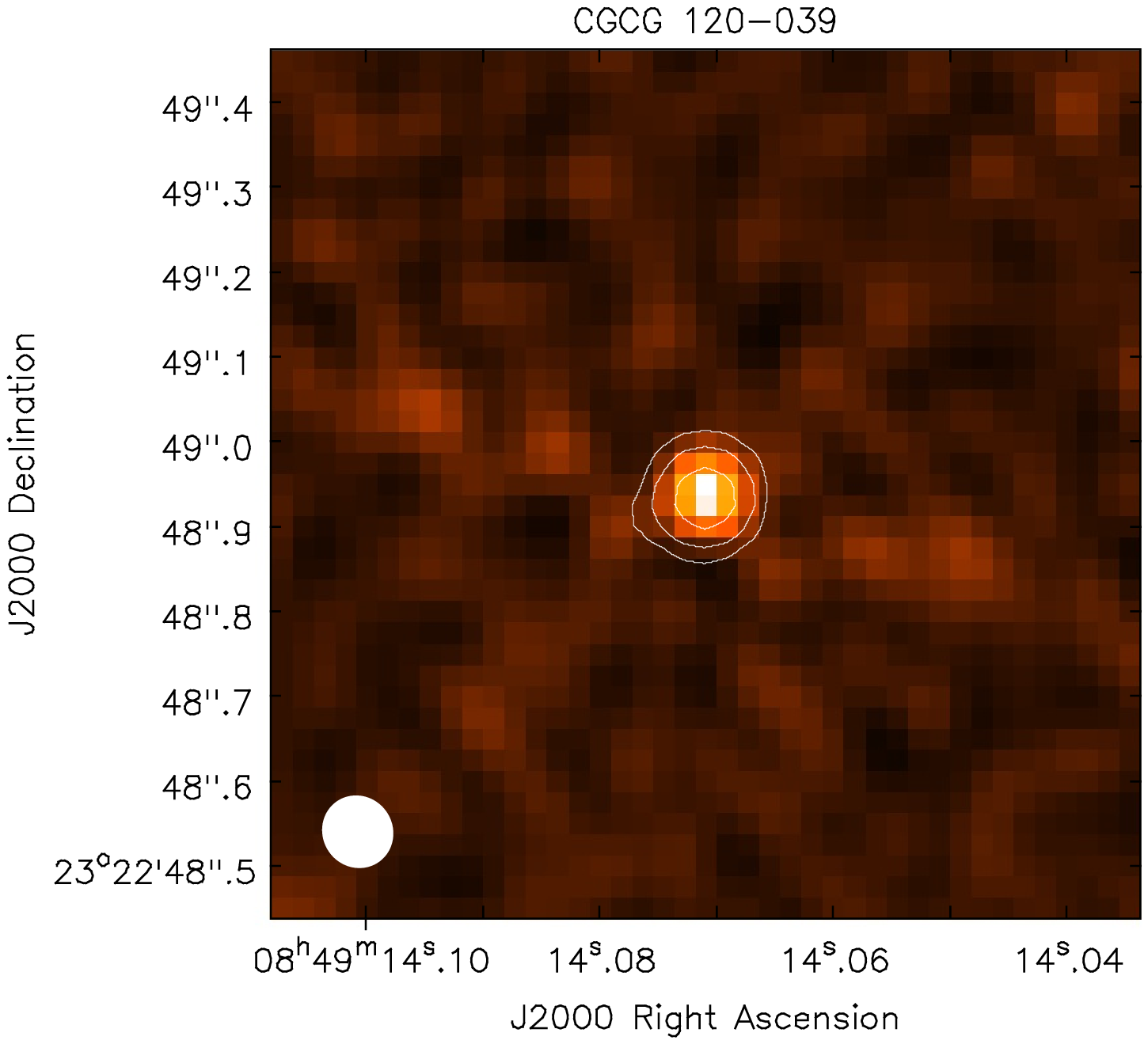}
\includegraphics[width=0.465\textwidth]{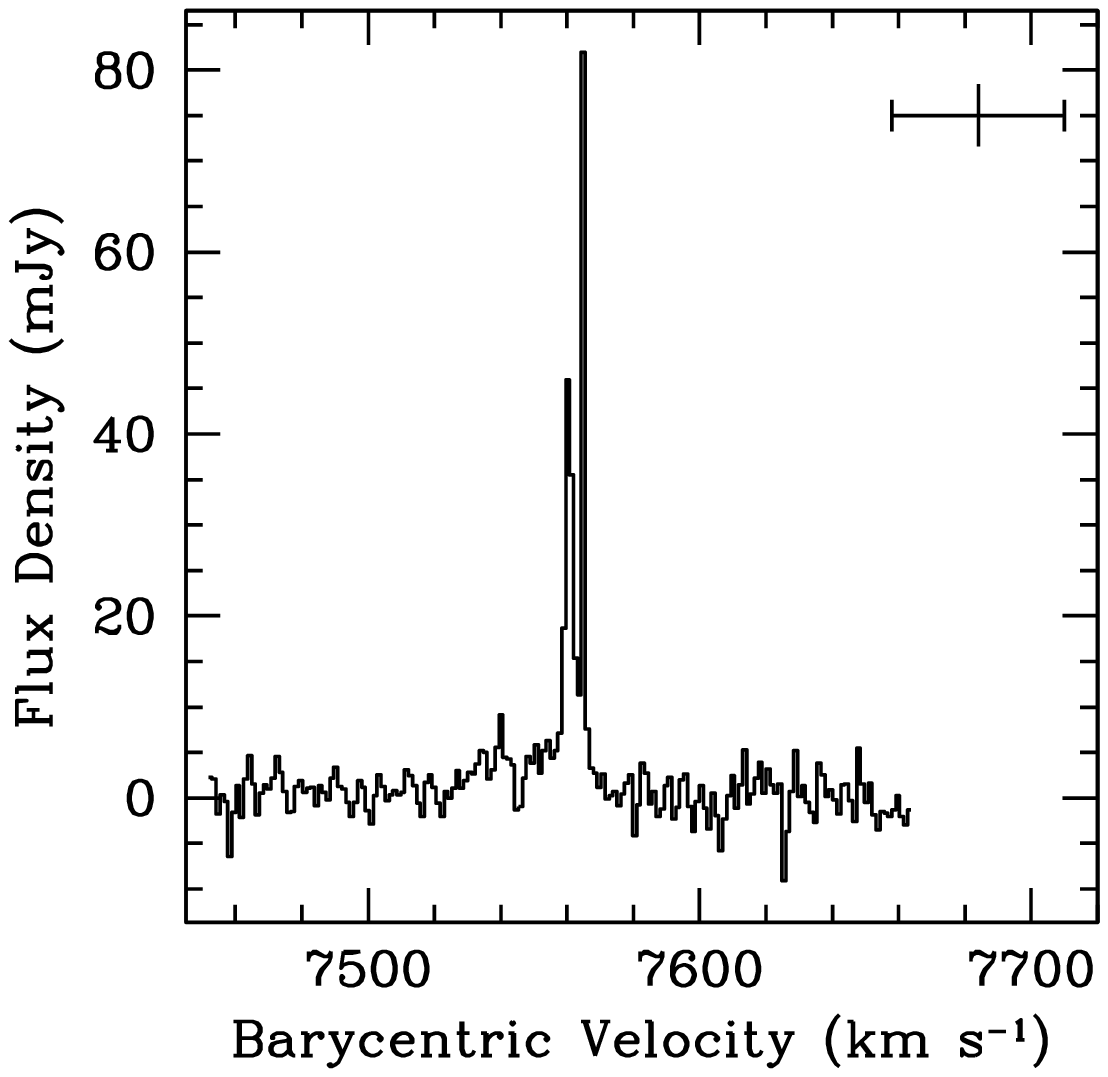}
\caption{Left:  CGCG 120$-$039 integrated water maser (image) and 20 GHz radio continuum  (contours) maps.  Continuum contours 
indicate 4, 8, and 16 times the rms noise listed in Table \ref{tab:obs}.   The spectral line beam is shown in the lower left
(properties listed in Table \ref{tab:obs}). The 1\arcsec\ field of view is equivalent to 519 pc.  
Right:  water maser spectrum with the systemic velocity and its uncertainty indicated by the vertical bars (Table \ref{tab:masers}).  
The spectrum is roughly centered on the previous single-dish water maser detection (see Section \ref{sec:discussion}).  
\label{fig:CGCG120-039}}
\end{figure*}

\begin{figure*}
\includegraphics[width=0.52\textwidth]{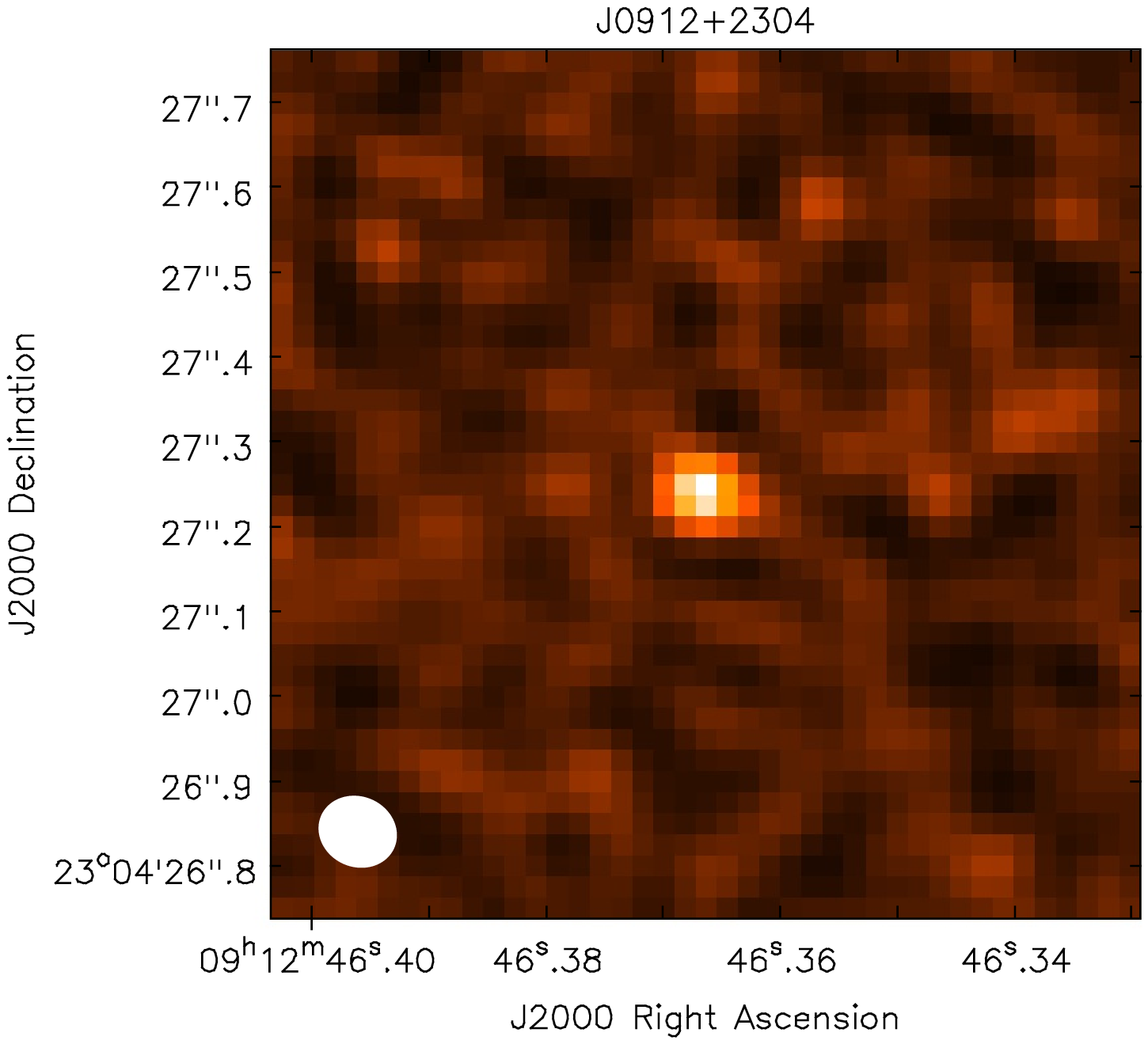}
\includegraphics[width=0.475\textwidth]{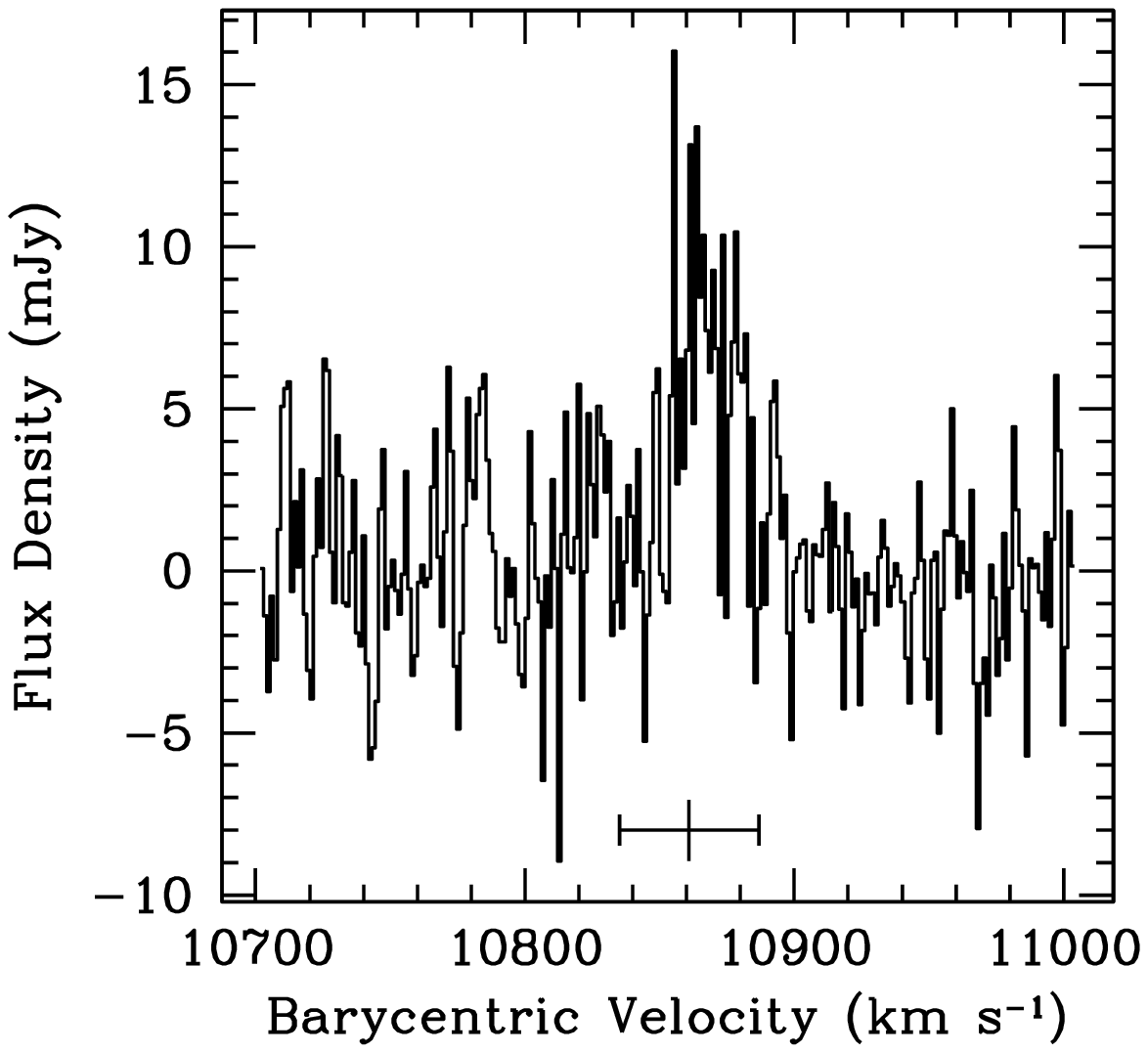}
\caption{Left:  J0912+2304  integrated water maser map.  The 20 GHz continuum was not significantly detected.
The spectral line beam is shown in the lower left
(properties listed in Table \ref{tab:obs}). The 1\arcsec\ field of view is equivalent to 723 pc.  
Right:  water maser spectrum with the systemic velocity and its uncertainty indicated by the vertical bars  (Table \ref{tab:masers}).  
The spectrum is centered on the previous single-dish water maser detection (see Section \ref{sec:discussion}).  
\label{fig:2MASX0912}}
\end{figure*}

\begin{figure*}
\includegraphics[width=0.53\textwidth]{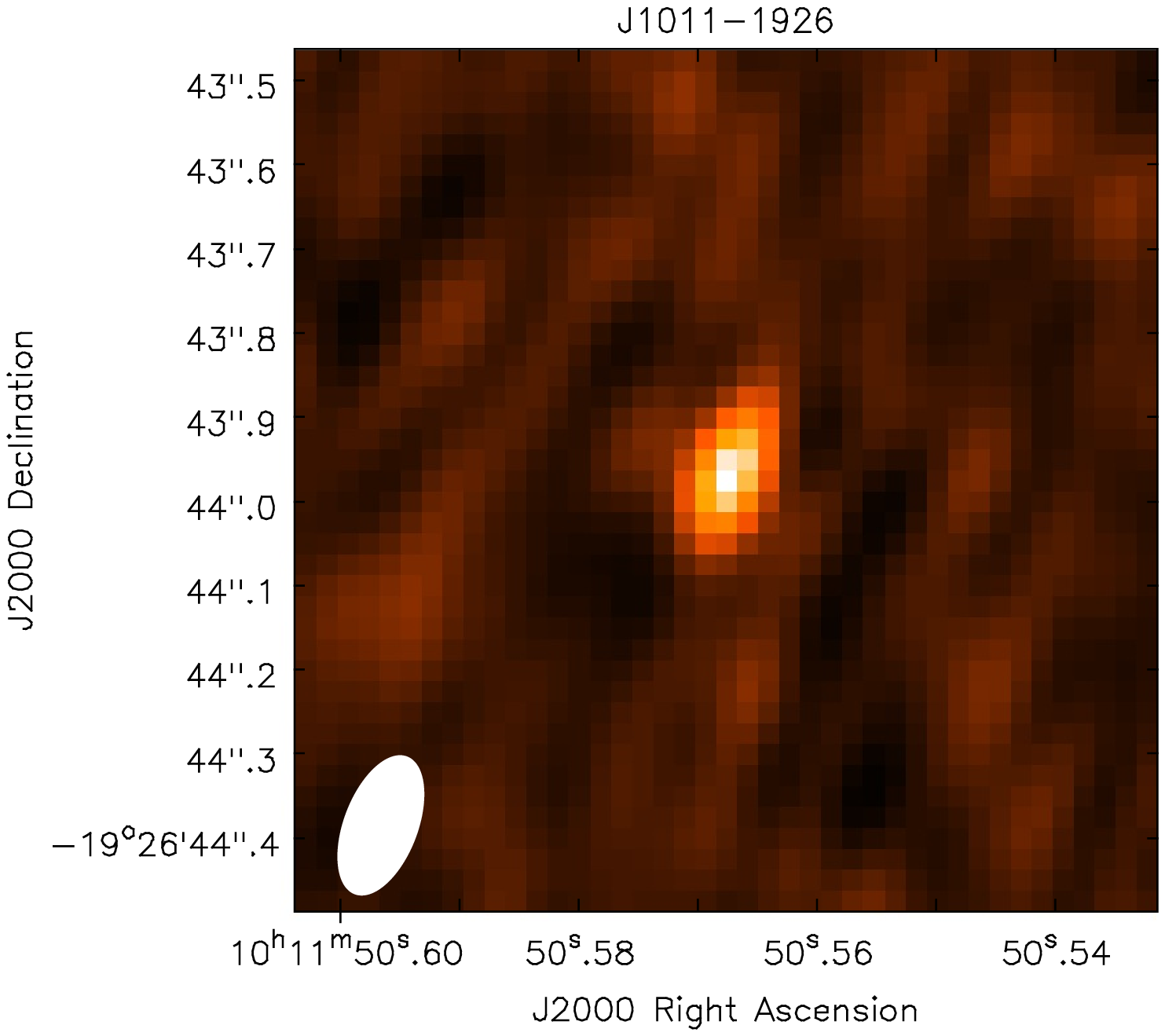}
\includegraphics[width=0.465\textwidth]{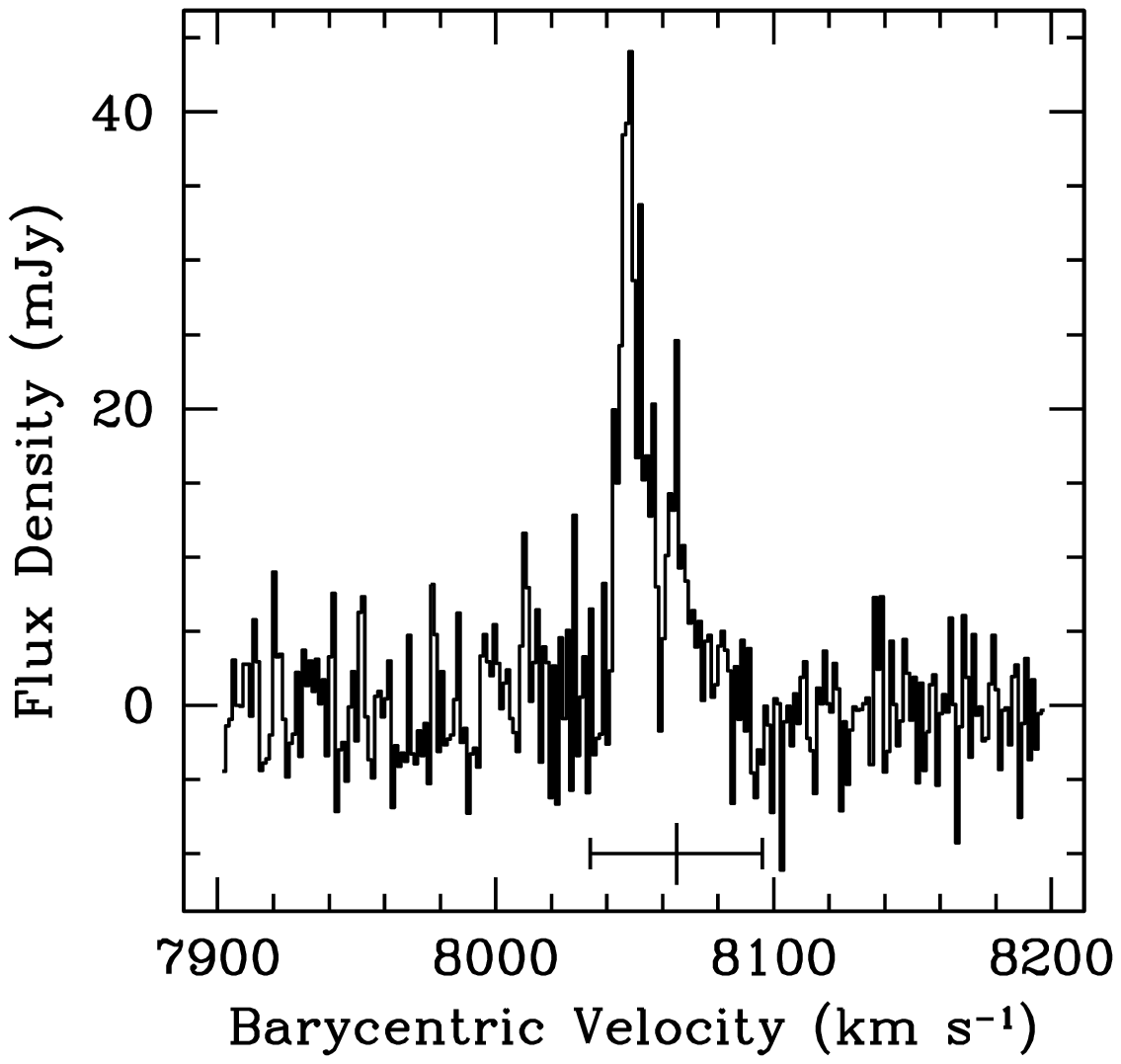}
\caption{Left: J1011$-$1926  integrated water maser map.  The 20 GHz continuum was not significantly detected.
The spectral line beam is shown in the lower left
(properties listed in Table \ref{tab:obs}). The 1\arcsec\ field of view is equivalent to 564 pc.  
Right:  water maser spectrum with the systemic velocity and its uncertainty indicated by the vertical bars (Table \ref{tab:masers}).  
The spectrum is centered on the previous single-dish water maser detection (see Section \ref{sec:discussion}).  
\label{fig:2MASX1011}}
\end{figure*}

\begin{figure*}
\includegraphics[width=0.51\textwidth]{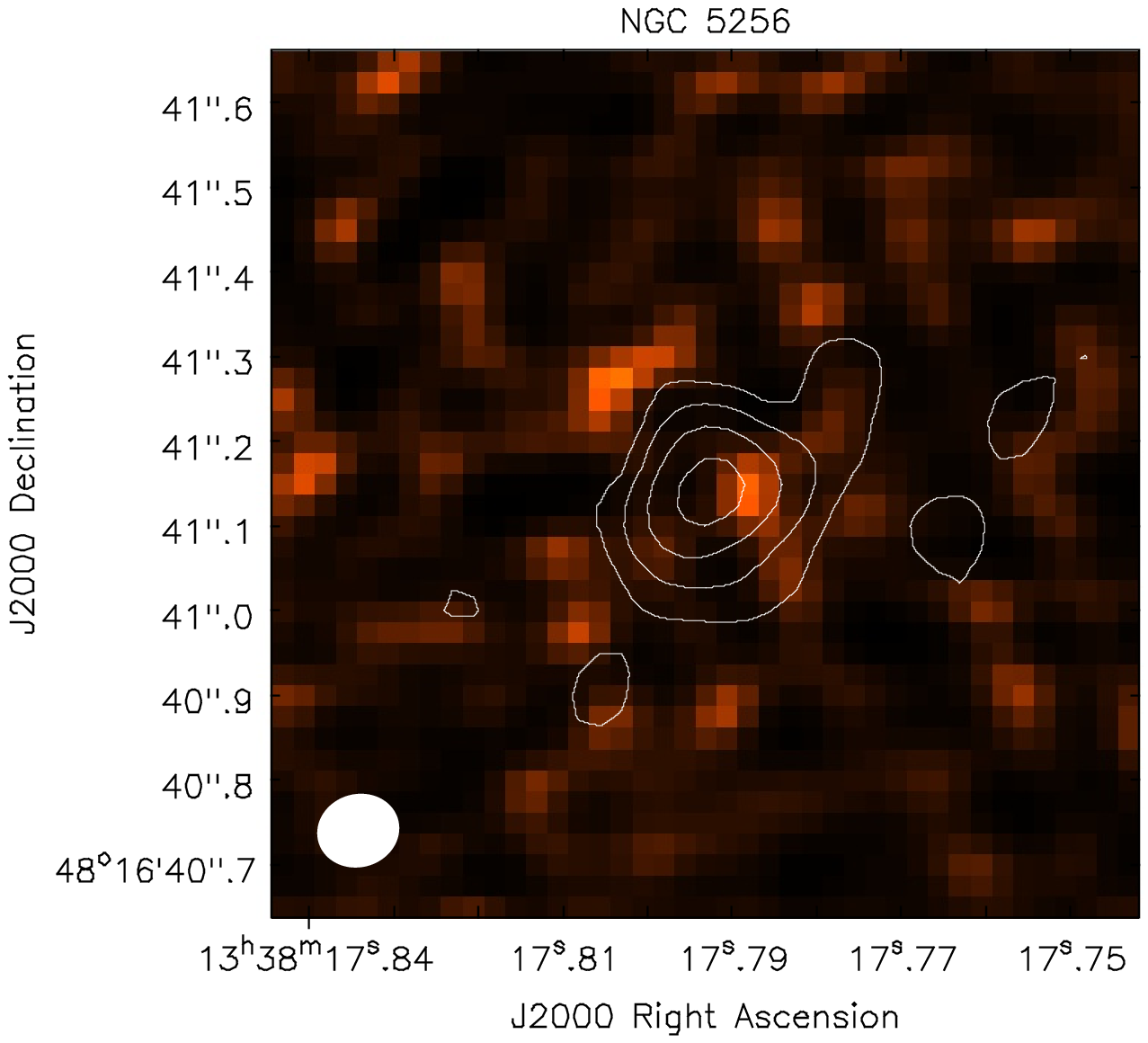}
\includegraphics[width=0.485\textwidth]{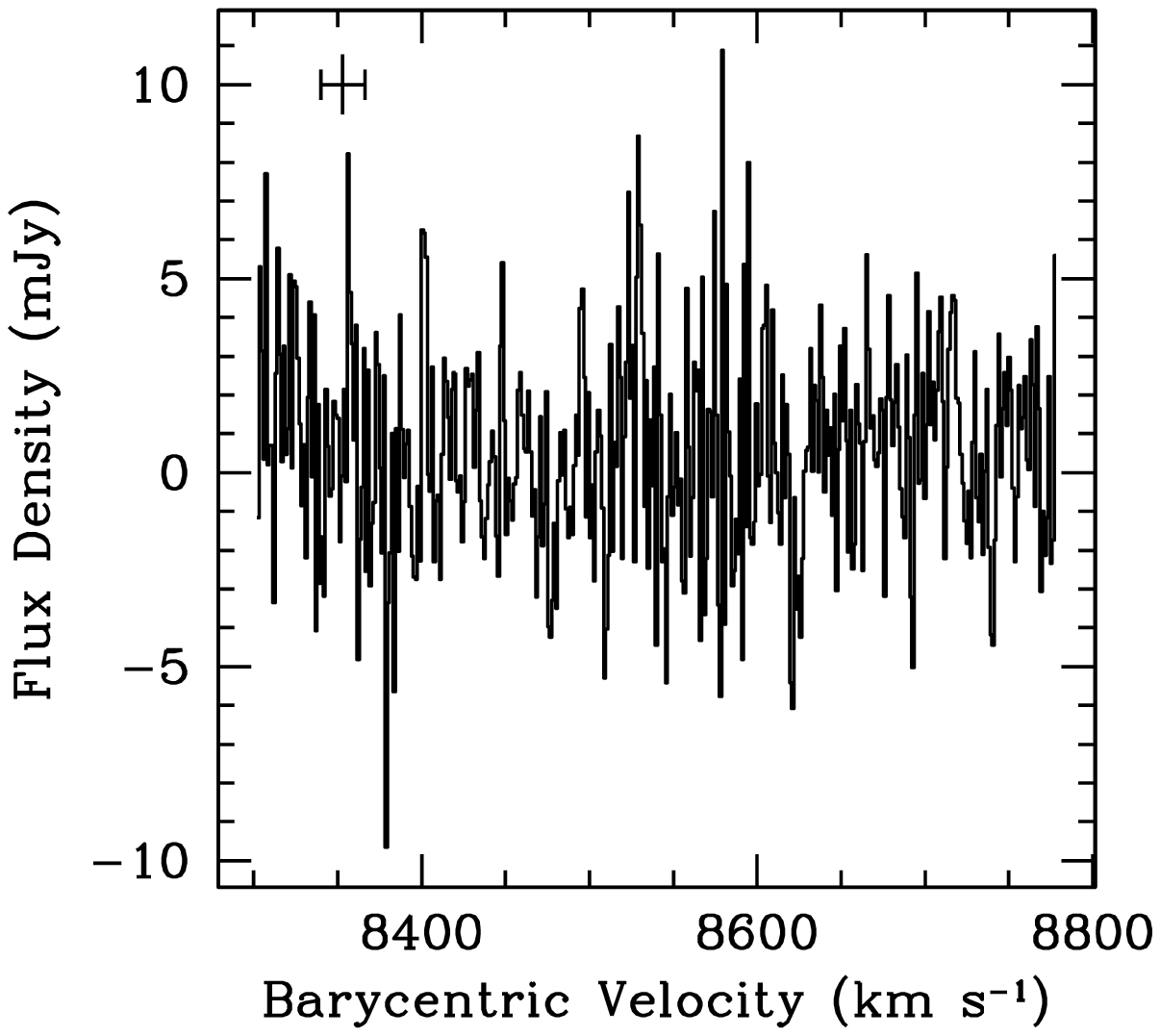}
\includegraphics[width=0.51\textwidth]{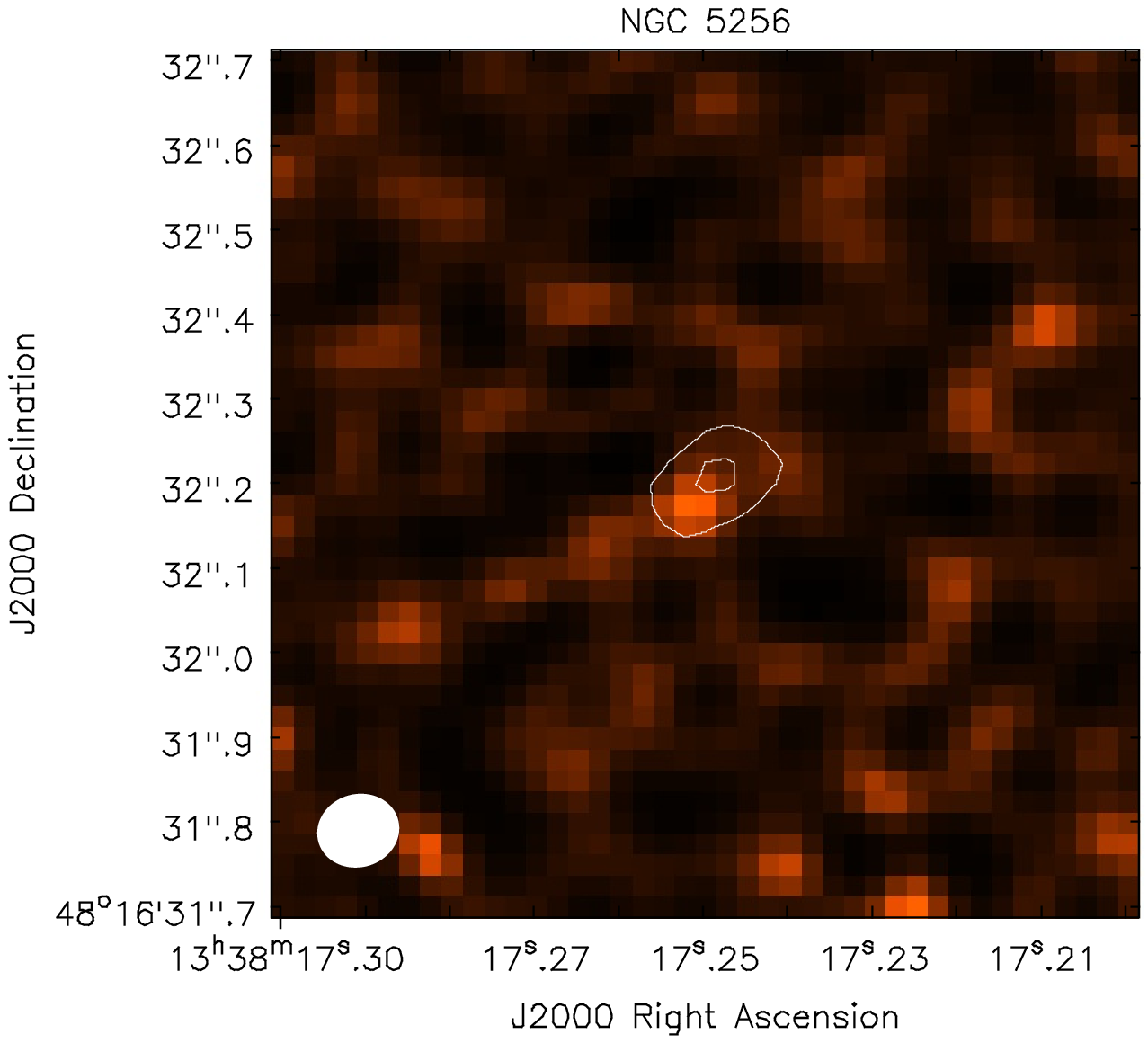}
\includegraphics[width=0.485\textwidth]{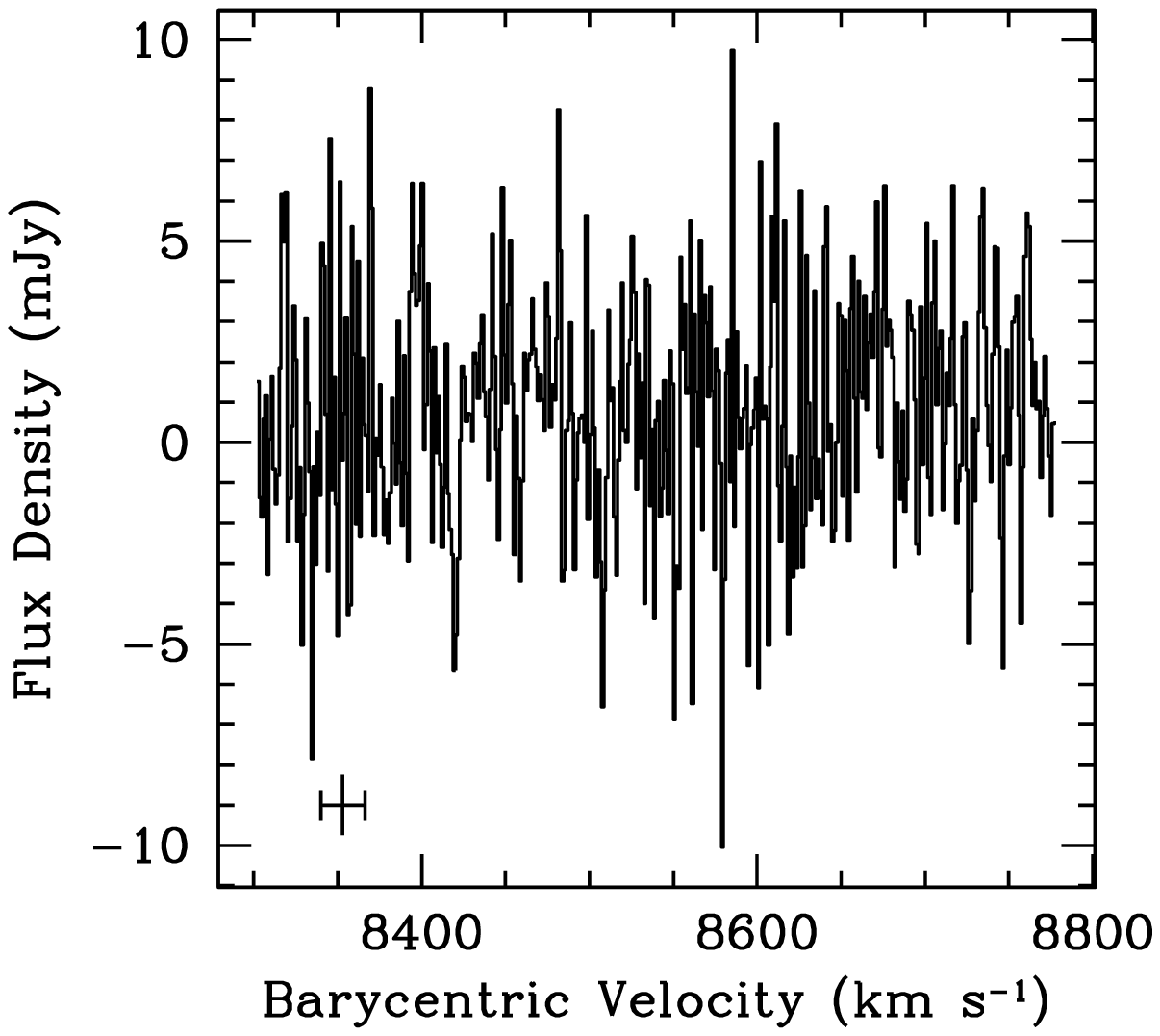}
\caption{Left:  NGC 5256 integrated water maser (image) and 20 GHz radio continuum  (contours) maps.  Continuum contours 
indicate 4, 8, 16, and 32 times the rms noise listed in Table \ref{tab:obs}.   The spectral line beam is shown in the lower left
(properties listed in Table \ref{tab:obs}). The 1\arcsec\ field of view is equivalent to 570 pc.  
Right:  water maser nondetection spectra at the continuum peaks with the systemic velocity and its uncertainty indicated by the vertical bars
\citep[$8353\pm13$ km s$^{-1}$;][]{RC3}.
The spectra are roughly centered on the previous single-dish water maser detection (see Section \ref{sec:discussion}).  
Two continuum sources were detected, but no water maser emission was detected toward either continuum source or in the 
larger field of view.
\label{fig:NGC5256}}
\end{figure*}

\begin{figure*}
\includegraphics[width=0.53\textwidth]{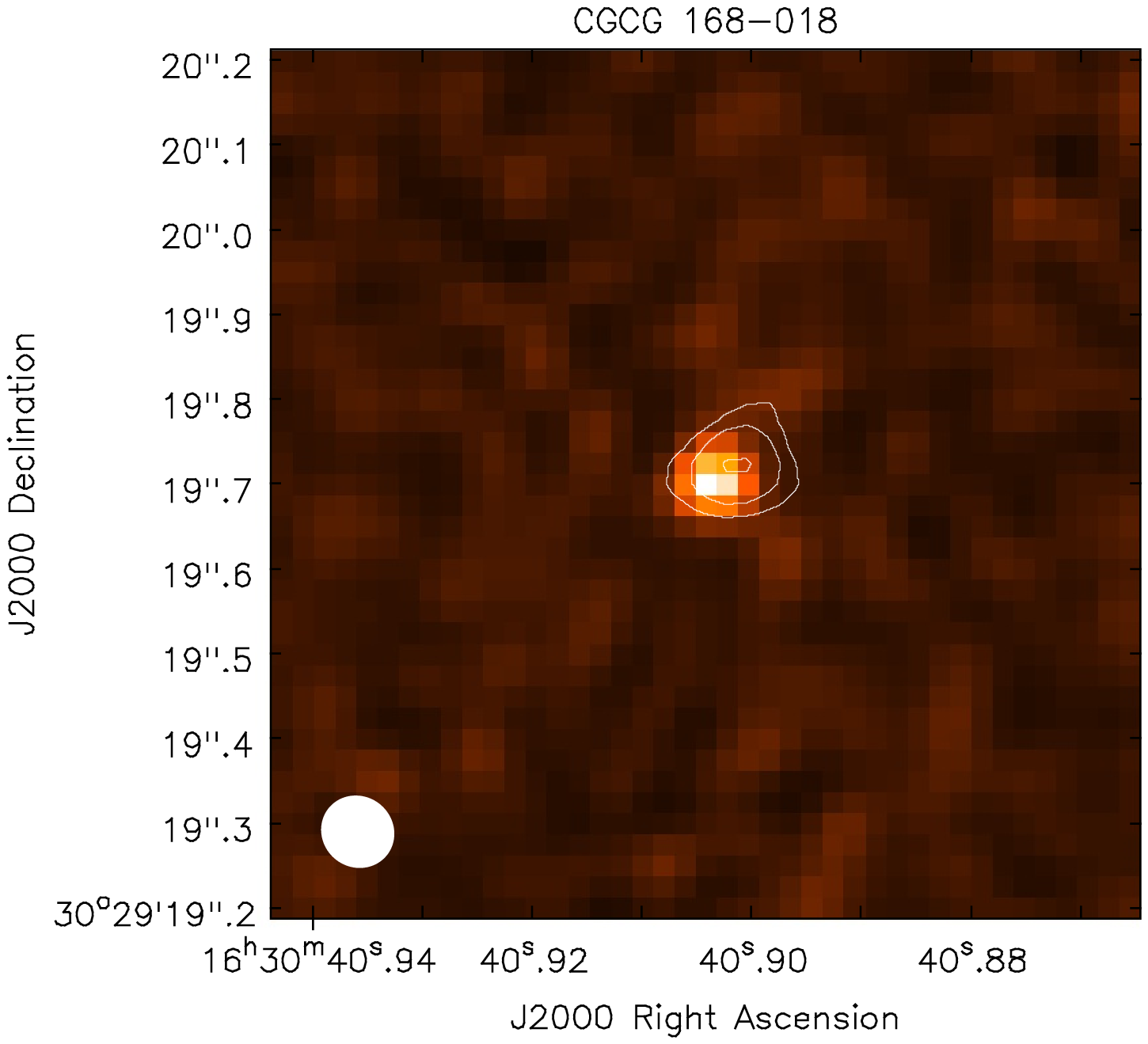}
\includegraphics[width=0.465\textwidth]{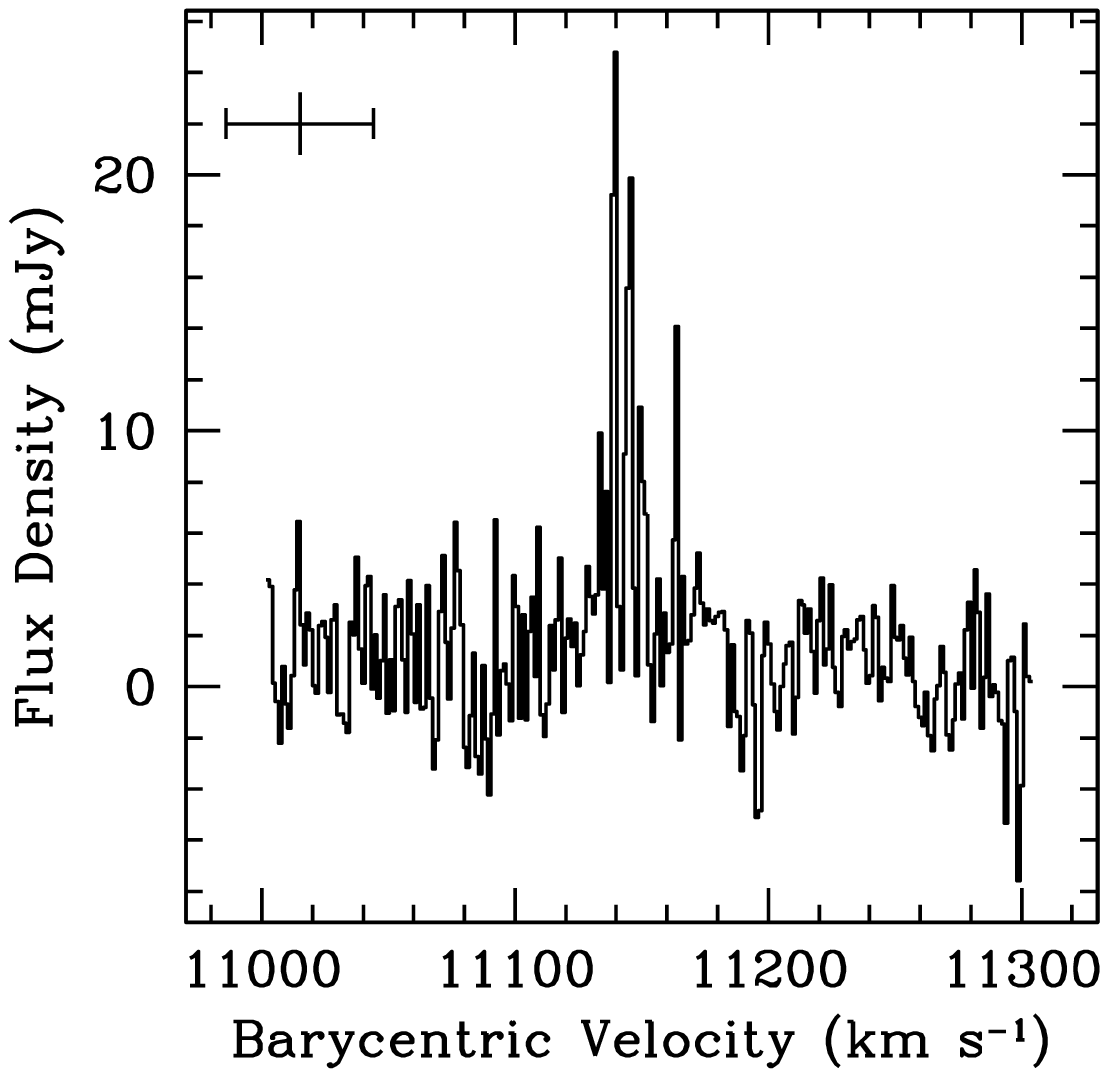}
\caption{Left:  CGCG 168$-$018 integrated water maser (image) and 20 GHz radio continuum  (contours) maps.  Continuum contours 
indicate 4, 8, and 16 times the rms noise listed in Table \ref{tab:obs}.   The spectral line beam is shown in the lower left
(properties listed in Table \ref{tab:obs}). The 1\arcsec\ field of view is equivalent to 737 pc.  
Right:  water maser spectrum with the systemic velocity and its uncertainty indicated by the vertical bars (Table \ref{tab:masers}).  
The spectrum is roughly centered on the previous single-dish water maser detection (see Section \ref{sec:discussion}).  
\label{fig:CGCG168-018}}
\end{figure*}

\begin{figure*}
\includegraphics[width=0.53\textwidth]{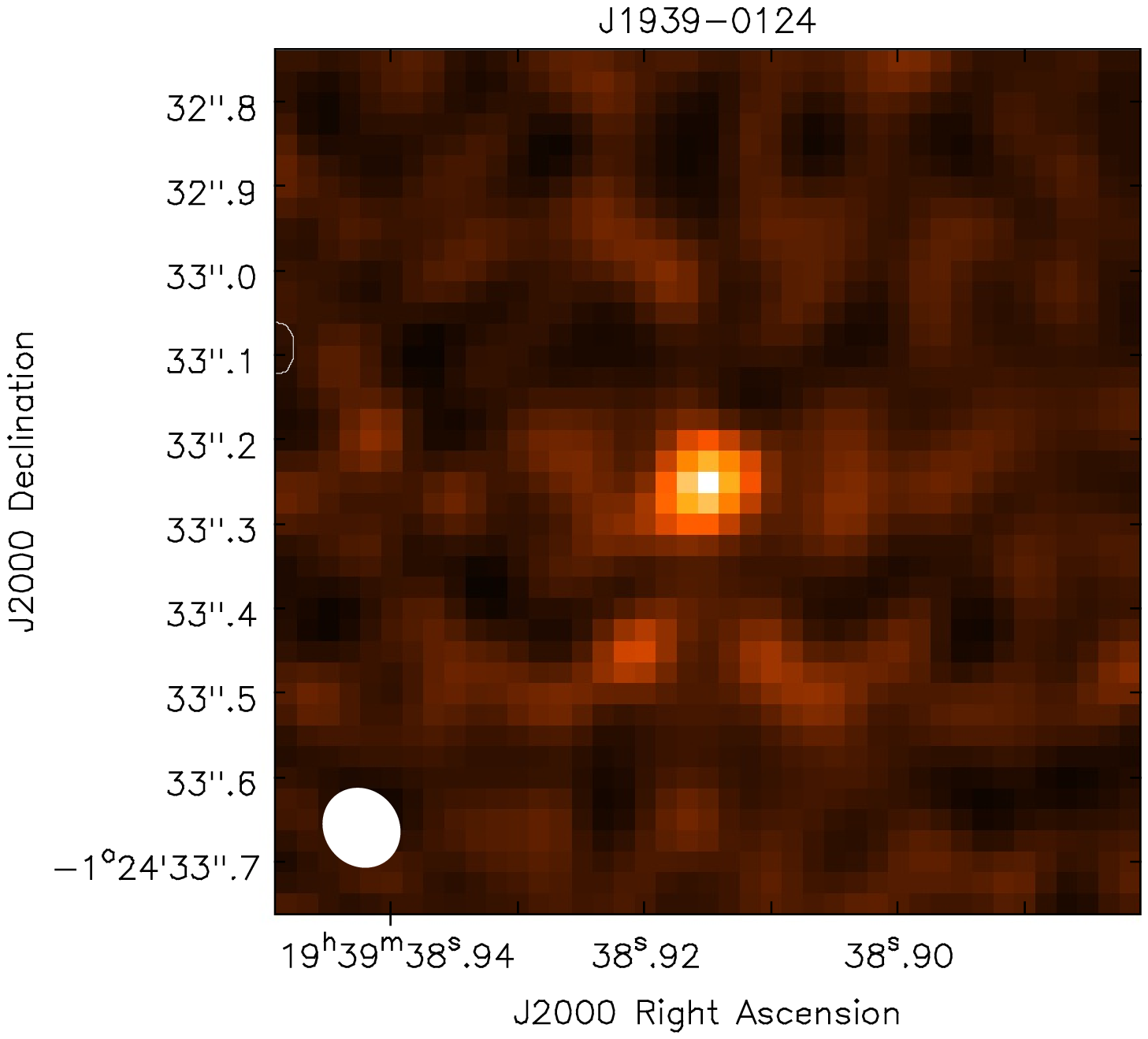}
\includegraphics[width=0.465\textwidth]{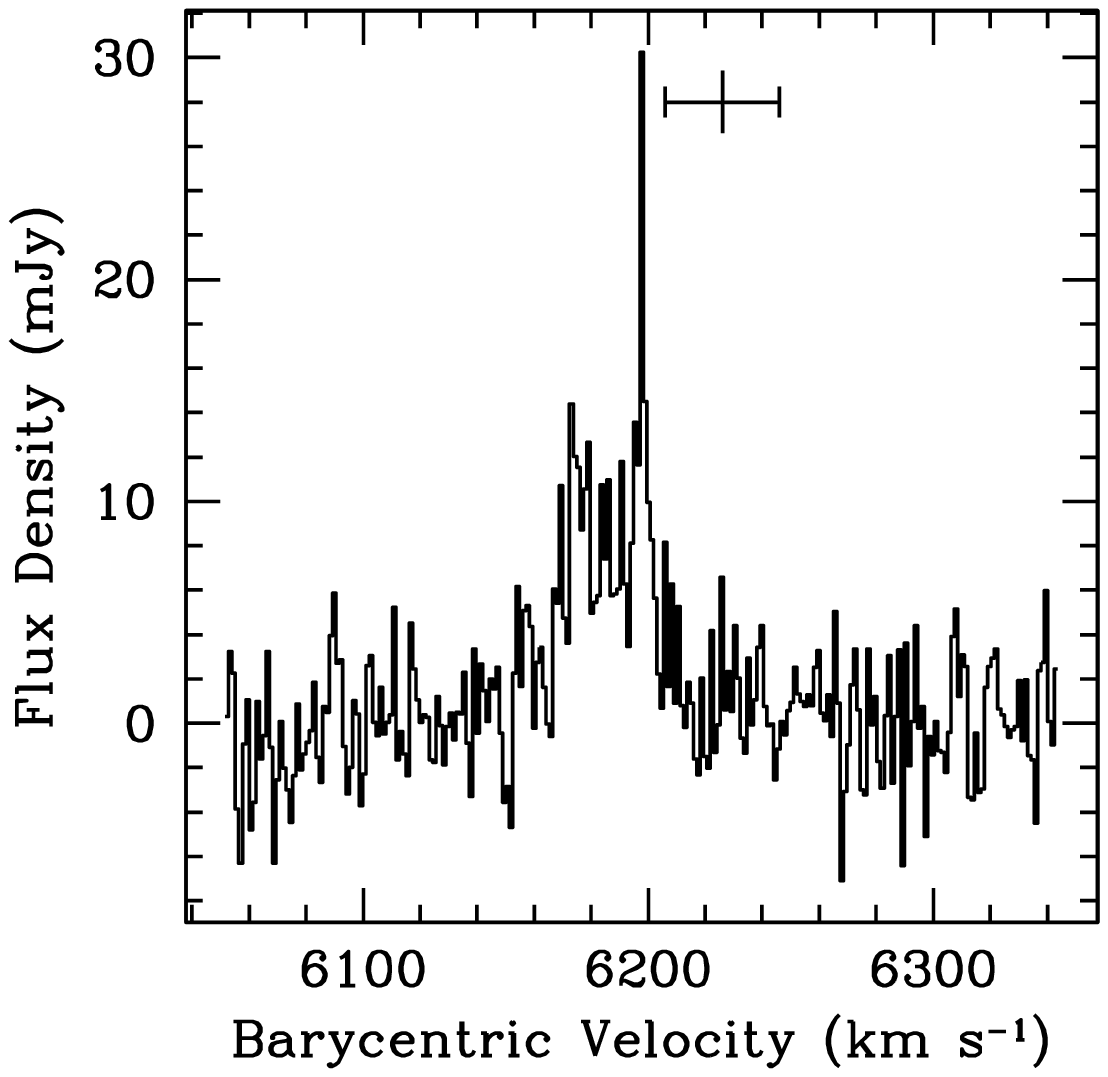}
\caption{Left:  J1939$-$0124  integrated water maser map.  The 20 GHz continuum was not significantly detected.
The spectral line beam is shown in the lower left
(properties listed in Table \ref{tab:obs}). The 1\arcsec\ field of view is equivalent to 432 pc.  
Right:  water maser spectrum with the systemic velocity and its uncertainty indicated by the vertical bars (Table \ref{tab:masers}).  
The spectrum is roughly centered on the previous single-dish water maser detection (see Section \ref{sec:discussion}).  
\label{fig:2MASX1939}}
\end{figure*}

Water masers were detected in 9 out of 13 objects, and 20 GHz continuum emission was detected in 7 out of 13 objects.  
Only five objects show both maser and continuum emission, and two objects were detected in neither line nor continuum.  
Figures \ref{fig:NGC291}--\ref{fig:2MASX1939} show a 1\arcsec\ square field of view of the first moment maser maps and
continuum contours and they show spectra of the detected objects (in line, continuum, or both).  Table \ref{tab:positions} lists the 
maser and continuum centroids based on two-dimensional Gaussian fits.  For the spectral lines, these fits were made
to the integrated line maps.  The maser emission was universally unresolved, but the continuum emission was formally resolved 
when deconvolved from the beam in all but two objects, IC 485 and CGCG 168$-$018.  

Table \ref{tab:positions} also lists the maser-continuum offsets in angular and physical units. Offsets for four of the five objects 
detected in both maser and continuum are non-significant with $1\sigma$ uncertainties ranging from about 1 to 40 pc.  The only object
showing a significant offset between the maser and continuum centroids is CGCG 168$-$018, with a $21.6\pm2.7$ pc offset (Figure \ref{fig:CGCG168-018}).  

Table \ref{tab:masers} lists the measured and derived water maser properties:  peak and integrated flux densities, luminosity distance, 
isotropic line luminosity, the range of velocities spanned by the line emission 3$\sigma$ above the noise, the velocity of peak emission, and 
the adopted systemic velocity.  For J0804+3607, we list redshifts rather than velocities.  The detected maser velocities are consistent with previous 
observations, although the masers can be substantially offset from the systemic velocities, which are obtained from optical and HI 21 cm 
lines.  It is unclear whether these velocity offsets are physical (i.e., due to different line-emitting regions genuinely having different velocities, as
is seen in shock-induced maser emission), due to obscuration (optical vs. radio lines), or due to measurement error, particularly 
in optical redshifts.  We therefore do not rely on the velocity offset between the maser emission and the adopted systemic  
velocity as a criterion for assessing the likelihood of a maser arising from an inclined disk.  
Isotropic maser luminosities range from kilomaser values ($3.05 \pm 0.26\ L_\odot$ in NGC 520b) to the exceptionally luminous, 
$L_{iso} = (1.8 \pm 0.1) \times 10^4\ L_\odot$ in J0804+3607 (Section \ref{sec:discussion}).  

Table \ref{tab:continuum} shows the 20 GHz radio continuum properties of the seven detected objects.  We include the peak flux density, 
the integrated flux density, the spectral index derived solely from the 18--22 GHz bandpass, and the deconvolved angular size.  
Among the six continuum sources with enough signal-to-noise to derive a significant spectral index, four are steep spectrum 
($\alpha = -1$ to $-2$) and two are flat ($\alpha \simeq -0.2 \pm 0.2$).  One of the latter, CGCG 120$-$039, is only marginally resolved.\
IC 485, which does not have a spectral index measurement, but which shows both maser and continuum emission, does not 
have a resolved continuum.

\floattable
\begin{deluxetable}{cllllcc}
\tablecaption{Maser and Radio Continuum Positions \label{tab:positions}}
\tablecolumns{7}
\tablewidth{0pt}
\tablehead{
\colhead{Galaxy} & 
\multicolumn{2}{c}{Maser Centroid\tablenotemark{a}} &
\multicolumn{2}{c}{Continuum Centroid\tablenotemark{b}} &
\multicolumn{2}{c}{Offset} \\
\colhead{} & 
\colhead{RA} & 
\colhead{Dec} & 
\colhead{RA} & 
\colhead{Dec} & 
\colhead{Angular} & 
\colhead{Physical}\\
\colhead{} & 
\colhead{(hms)} & 
\colhead{(dms)} & 
\colhead{(hms)} & 
\colhead{(dms)} & 
\colhead{(milliarcsec)} & 
\colhead{(pc)} 
}
\startdata
NGC 291          & \nodata & \nodata & 00:53:29.9101(11) & $-$08.46.03.740(14) & \nodata & \nodata \\ 
NGC 520b        & 01:24:34.91412(14) & +03.47.29.7864(22) & 01:24:34.9099(28) & +03.47.29.7783(92) & 64(42) & 8.6(5.6) \\ 
J0350$-$0127 & 03:50:00.352168(29) & $-$01.27.57.39574(53) & \nodata & \nodata & \nodata & \nodata \\ 
IC 485              & 08:00:19.752486(74) & +26.42.05.0526(10) & 08:00:19.7515(14) & +26.42.05.050(14) & 13(19) & 7.6(10.5)\\ 
J0804+3607 &  08:04:31.01144(40) & +36.07.18.1937(52) & 08:04:31.01138(12) & +36.07.18.19927(91) & 5.6(5.3) & 39.6(37.2) \\ 
CGCG 120$-$039 & 08:49:14.07078(16) & +23.22.48.9408(20) &  08:49:14.07097(13) & +23.22.48.9346(17) & 6.7(2.7) & 3.5(1.4) \\ 
J0912+2304    & 09:12:46.36659(33) & +23.04.27.2421(28) & \nodata & \nodata & \nodata & \nodata \\ 
J1011$-$1926 & 10:11:50.56731(17) & $-$19.26.43.9645(59) & \nodata & \nodata & \nodata & \nodata \\ 
UGC 7016       & \nodata & \nodata & \nodata & \nodata & \nodata & \nodata \\ 
NGC 5256\tablenotemark{c} & \nodata & \nodata & 13:38:17.79219(15) & +48.16.41.1389(19) & \nodata & \nodata \\ 
                       & \nodata & \nodata & 13:38:17.24843(76) & +48.16.32.2095(92) & \nodata & \nodata \\ 
NGC 5691       & \nodata & \nodata & \nodata & \nodata & \nodata & \nodata \\ 
CGCG 168$-$018 &  16:30:40.90329(19) & +30.29.19.7066(29) &  16:30:40.90134(21) & +30.29.19.7216(20) & 29.3(3.6) & 21.6(2.7) \\ 
J1939$-$0124 & 19:39:38.91545(38) & $-$01.24.33.2553(39) & \nodata & \nodata & \nodata & \nodata \\ 
\enddata
\tablenotetext{a}{All detected maser emission was unresolved.  See Table \ref{tab:obs} for beam sizes.}
\tablenotetext{b}{All detected continuum emission was resolved except for IC 485 and CGCG 168$-$018.  
  See Table \ref{tab:continuum} for continuum measurements.}
\tablenotetext{c}{NGC 5256 shows two widely-separated continuum components (both are listed).}
\tablecomments{Coordinates are epoch J2000, and parenthetical values indicate uncertainties in the ultimate digits.}
\end{deluxetable}

\floattable
\begin{deluxetable}{crrrrcrrc}
\tablecaption{Water Maser Properties and Redshifts \label{tab:masers}}
\tablecolumns{9}
\tablewidth{0pt}
\tablehead{
\colhead{Galaxy} & 
\colhead{$S_{\rm peak}$\tablenotemark{a}} &
\colhead{$S_{\rm int}$\tablenotemark{b}} &
\colhead{$D_L$} & 
\colhead{$L_{\rm iso}$\tablenotemark{c}} & 
\colhead{$v_{\rm range}$\tablenotemark{d}} & 
\colhead{$v_{\rm peak}$} & 
\colhead{$v_{\rm sys}$} & 
\colhead{Ref.\tablenotemark{e}} \\
\colhead{} & 
\colhead{(mJy)} & 
\colhead{(mJy km s$^{-1}$)} & 
\colhead{(Mpc)} & 
\colhead{($L_\odot$)} & 
\colhead{(km s$^{-1}$)} & 
\colhead{(km s$^{-1}$)} & 
\colhead{(km s$^{-1}$)} & 
\colhead{} 
}
\startdata
NGC 520b        & 35(3) & 166(14) & 28 & 3.05(26) & 2270--2273 & 2272(1) & 2288(8) & 1 \\ 
J0350$-$0127 &  347(4) & 7200(150)  & 180 & 5181(108) & 12336--12393\tablenotemark{f} & 12369(1) & 12322(18) & 2 \\ 
IC 485              & 78(2) & 2470(130) & 125 & 868(46) & 8307--8387 & 8356(1) & 8338(10) & 3 \\  
J0804+3607\tablenotemark{g} & 9(1) & 80(6) & 3997 & $1.8(1)\times10^4$ & 0.66038--0.66051 & 0.66045(1) & 0.65654(37)  & 4 \\ 
CGCG 120$-$039 & 82(2) & 438(46) & 116 &  133(14) & 7559--7565 & 7565(1) & 7684(26) & 5\\  
J0912+2304    & 16(3) & 252(44) & 164 & 151(26) & 10855--10878 & 10855(1) & 10861(26) &  6\\ 
J1011$-$1926 & 44(4) & 624(82) & 123 & 213(28) & 8043--8065 & 8048(1) & 8065(31) & 7 \\ 
CGCG 168$-$018 & 25(2) & 279(36) & 162 & 163(21) & 11134--11164 & 11140(1) & 11015(29) & 5 \\  
J1939$-$0124 & 30(2) & 519(80) & 93 & 102(16) & 6170--6206 & 6198(1) & 6226(20)  & 8 \\ 
\enddata
\tablenotetext{a}{The peak flux density was obtained from a spectrum formed from a single beam centered 
on the peak (unresolved) maser emission.}
\tablenotetext{b}{The integrated flux density of the water maser complex was 
obtained from fitting a single two-dimensional Gaussian to the velocity-integrated spectral line cube.}
\tablenotetext{c}{The isotropic luminosity is computed from the integrated line flux density $S_{int}$ via 
  $L_{iso} = 23.1\ L_\odot \times S_{\rm int} ({\rm mJy\ km\ s}^{-1}) \times D_L ({\rm Gpc})^2/(1+z)$, where $D_L$ is the luminosity distance and 
  $z$ is the cosmological redshift.}
\tablenotetext{d}{$v_{\rm range}$ is the velocity range over which the water maser exceeds 3$\sigma$ significance.}
\tablenotetext{e}{References for the systemic velocities:
1 -- \citet{springob2005}; 
2 -- \citet{huchra2012};
3 -- \citet{RC3};
4 -- \citet{hewett2010};
5 -- \citet{SDSSDR5};
6 -- \citet{SDSSDR6}; 
7 -- \citet{6dFDR3};
8 -- \citet{theureau2007}.}
\tablenotetext{f}{There appears to be an additional narrow maser line at 12422 km s$^{-1}$ (Figure \ref{fig:2MASX0350}).}
\tablenotetext{g}{We list redshifts instead of velocities for $v_{\rm range}$, $v_{\rm peak}$, and $v_{\rm sys}$ for J0804+3607.}
\tablecomments{Only objects with detected water maser emission are listed (see Table \ref{tab:obs} for rms noise values for nondetections).
Parenthetical values indicate uncertainties in the ultimate digits.}
\end{deluxetable}

\floattable
\begin{deluxetable}{ccccCcC}
\tablecaption{20 GHz Radio Continuum Properties \label{tab:continuum}}
\tablecolumns{7}
\tablewidth{0pt}
\tablehead{
\colhead{Galaxy} & 
\colhead{$S_{\rm peak}$\tablenotemark{a}} &
\colhead{$S_{\rm int}$\tablenotemark{a}} &
\colhead{$\alpha$\tablenotemark{b}} &
\multicolumn{3}{c}{Size\tablenotemark{a}} \\
\cline{5-7}
\colhead{} & 
\colhead{} & 
\colhead{} & 
\colhead{} & 
\colhead{Angular} & 
\colhead{PA} & 
\colhead{Physical}\\
\colhead{} & 
\colhead{($\mu$Jy bm$^{-1}$)} & 
\colhead{(mJy)} & 
\colhead{} & 
\colhead{(milliarcsec)} & 
\colhead{($^\circ$)} & 
\colhead{(pc)} 
}
\startdata
NGC 291          & 263(24) & 1.93(20) & $-$2.0(3) & 451(52)\times146(19) & 51.8(3.4) & 177(20)\times57(7) \\ 
NGC 520b        & 272(31) & 4.31(52) & $-$0.23(19) & 775(100)\times164(24) & 95.6(2.3) & 105(14)\times22(3) \\ 
IC 485              &  77(15)  &0.180(48)& \nodata & < 93\times88            & $-$69.5 &    < 52\times49 \\ 
J0804+3607 & 4247(83) & 4.71(16) & $-$1.5(3) & 62(12)\times23(13) & 71(17) & 437(85)\times162(92) \\ 
CGCG 120$-$039 & 482(20) & 0.555(39) & $-$0.27(16) & 44(8)\times23(14) & 133(33) & 23(4)\times12(7) \\ 
NGC 5256\tablenotemark{c} & 633(28) & 1.543(68) & $-$2.0(2) & 127(6)\times96(8) & 148.7(9.6) & 72(3)\times55(5)\\ 
                      & 78(17) & 0.48(11) & \nodata & 433(23)\times68(58) & 163.1(3.1) & 247(13)\times39(33) \\ 
CGCG 168$-$018 &  267(16) & 0.308(30) & $-$0.95(17) & < 87\times85 & 77.8 & < 64\times63\\ 
\enddata
\tablenotetext{a}{The peak flux density, integrated flux density, and the radio source size and orientation were 
obtained from fitting a single two-dimensional Gaussian to the source image and 
deconvolving the source from the beam. Upper limits indicate unresolved detections and list the continuum beam parameters.}
\tablenotetext{b}{The spectral index $\alpha$ was measured at the peak of the continuum emission solely from within the 18--22 GHz bandpass
and follows the convention $S_\nu \propto\nu^\alpha$. It is not listed for objects with inadequate signal-to-noise in the 20 GHz continuum. }
\tablenotetext{c}{NGC 5256 shows two widely-separated continuum components (both are listed).}
\tablecomments{Only objects with detected 20 GHz continuum emission are listed (see Table \ref{tab:obs} for rms noise values for nondetections).  
Parenthetical values indicate uncertainties in the ultimate digits.}
\end{deluxetable}

\section{Analysis} \label{sec:analysis}

\begin{figure}
\includegraphics[width=0.45\textwidth]{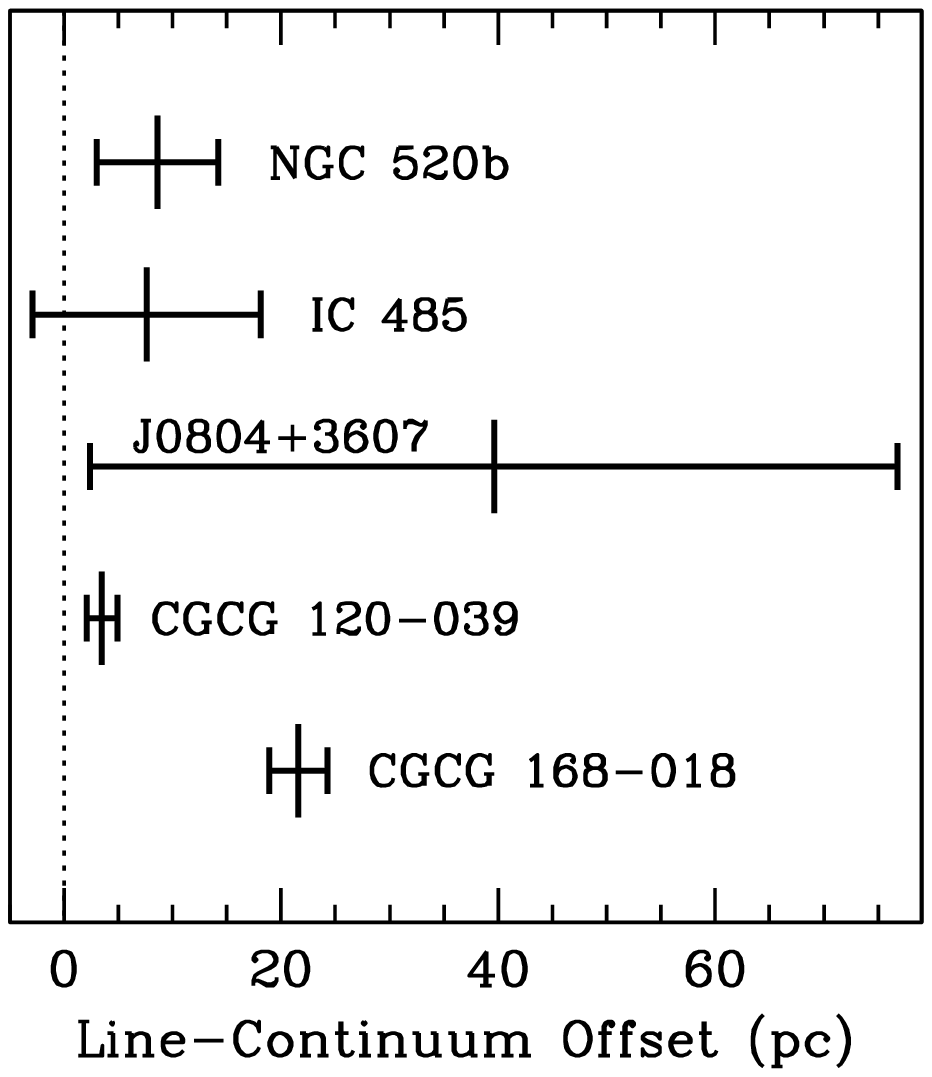}
\caption{Maser-continuum offsets for those objects detected in both.  Only CGCG 168$-$018 shows a significant offset.  Error bars are 
plotted symmetrically.
\label{fig:offsets}}
\end{figure}

Figure \ref{fig:offsets} shows the projected physical offset between the water maser line and continuum centroids for the 
five objects detected in both line and continuum (Table \ref{tab:positions}).  Only CGCG 168$-$018 shows a significant offset
of $29.3\pm3.6$ milliarcseconds or $21.6\pm2.7$ pc, but based 
on its 18--22 GHz  continuum spectral index of $-0.95\pm0.17$ (Table \ref{tab:continuum}), 
the continuum could be jet emission and not the core of the AGN.  
All other maser-continuum
offsets are not significant and are therefore consistent with
inclined maser disk expectations. 

\section{Discussion}\label{sec:discussion}

Among the 13 water maser hosts, we did not detect water masers in four objects:
NGC 291, UGC 7016, NGC 5256, and NGC 5691.
The water maser in NGC 520b is likely associated with star formation and is rejected 
as an inclined maser disk candidate (see \ref{subsubsec:NGC520b}).  
Objects with unresolved masers but not detected in the 20 GHz continuum remain ambiguous 
(these are J0350$-$0127, J0912+2304, J1011$-$1926, and J1939$-$0124; note that none of these
have other sub-arcsecond radio continuum observations), as does 
CGCG 168$-$018, the only object to show a significant maser-continuum offset (see \ref{subsubsec:CGCG168-018}).
IC 485, J0804+3607, and CGCG 120$-$039 remain inclined maser disk candidates:  they show no 
significant maser-continuum offset and no high velocity lines (by selection).  While the maser
emission from IC 485 is broad and multi-component, this type of structure spanning $\sim$100 km s$^{-1}$ is seen in the systemic masers
in many disk systems \citep[e.g.,][]{kuo2011}.  J0808+3607 and CGCG 120$-$039 have narrow maser lines and are particularly 
good inclined disk candidates.  

\subsection{Individual Objects}
 
We discuss the previous water maser observations, general characteristics, and our results for each 
individual object below.

\subsubsection{NGC 291}   
NGC 291 is a barred spiral galaxy with a Seyfert 2 nucleus \citep{kewley2001,nair2010}.
The maser was detected as a narrow $\sim$60 mJy line in 2006 by GBT program 
GBT06A-009,\textsuperscript{\ref{footnote:maser_url}}
but we did not detect it (2.6 mJy beam$^{-1}$ rms per 1.2 km s$^{-1}$ 
channel; Table \ref{tab:obs} and Figure \ref{fig:NGC291}) nor did \citet{kondratko2006} in 2002 (15 mJy rms per 1.3 km s$^{-1}$ 
channel).  
We detect extended 20 GHz continuum with 18--22 GHz spectral index $-2.0\pm0.3$
(Table \ref{tab:continuum}).

\subsubsection{NGC 520b}    \label{subsubsec:NGC520b}
NGC 520b is one galaxy in the colliding pair in NGC 520 \citep[e.g.,][]{stanford1990}.  
\citet{castangia2008} detected a maser in 2005, measuring a $\sim40$ mJy peak with the VLA 
and $L_{iso} \sim 1\  L_\odot$.  Our VLA detection shows a $35\pm3$ mJy peak and
$L_{iso} = 3.05\pm0.26\ L_\odot$ (Figure \ref{fig:NGC520b} and Table \ref{tab:masers}).  
Our VLA map shows extended 20 GHz radio continuum emission (Figure \ref{fig:NGC520b}, bottom) with a 
flat 18--22 GHz spectral index of $-0.23\pm0.19$ (Table \ref{tab:continuum}) and  a similar east-west 
morphology to the \citet{castangia2008} 14.9 GHz map.  In our map the maser is slightly north of the \citet{castangia2008}
maser position.  There is a significant maser-continuum peak offset, although the single-component centroid 
fits listed in Table \ref{tab:positions} are consistent because the radio emission is extended.  \citet{castangia2008}
favor a star formation origin for the water maser in NGC 520b, but cannot rule out a low luminosity AGN.  
The preponderance of evidence suggests that NGC 520b is a poor candidate for an inclined maser disk.

\subsubsection{J0350$-$0127}  
J0350$-$0127 is an almost otherwise unknown spiral or possibly irregular galaxy included in the 2MASS Redshift Survey \citep{huchra2012}
and detected in water maser emission by the GBT programs GBT09-051\textsuperscript{\ref{footnote:maser_url}} in 2010 and 
GBT10C-019\textsuperscript{\ref{footnote:maser_url}} in 2011.  
The peak flux density was $\sim$350 mJy in 2011, in good agreement with our VLA measurement of $347\pm4$ mJy (Table \ref{tab:masers} and 
Figure \ref{fig:2MASX0350}).  
This is a broad and luminous maser ($L_{\rm iso} = 5181\pm108\ L_\odot$) with no associated 20 GHz continuum, down to an rms noise of 15 $\mu$Jy beam$^{-1}$
(Table \ref{tab:obs}).  
Without a continuum detection, the nature of this maser remains ambiguous, although it is almost certainly associated with an AGN.

\subsubsection{IC 485}  
IC 485 is a spiral galaxy detected in the water maser line in 2006 in GBT program
GBT06C-035.\textsuperscript{\ref{footnote:maser_url}}  The maser has a 
$\sim$80 mJy peak and shows a broad profile.  \citet{zhu2011} lists this object as 
a maser nondetection.  Our VLA observations show a 
broad, multi-component maser with a similar peak ($78\pm2$ mJy) 
and a high luminosity, $L_{iso} = 868\pm46\ L_\odot$ (Figure \ref{fig:IC485} and Table \ref{tab:masers}).  The 20 GHz continuum 
is detected but unresolved and faint  ($77\pm15$ $\mu$Jy beam$^{-1}$ peak flux density; Table \ref{tab:continuum}).
This galaxy was also detected at 1.4 GHz (4.4 mJy), and its dominant radio energy source was classified as star formation 
by \citet{condon2002}.  This classification does not exclude the presence of an AGN: \citet{liu2011} classify the optical nucleus 
of IC 485 as a LINER.  
The maser-continuum offset is not significant --- less than 10.5 pc (1$\sigma$; Table \ref{tab:positions}) --- so this maser remains an inclined disk candidate.

\subsubsection{J0804+3607}   
This is a type 2 quasar showing luminous water maser emission at $z\simeq0.66$ \citep{zakamska2003,barvainis2005}.  
The isotropic maser luminosity was $2.31\pm0.46\times10^4\ L_\odot$ in 2005 
\citep[][error from the quoted 20\% calibration uncertainty]{barvainis2005} and $1.8\pm0.1\times10^4\ L_\odot$ in 2015 (this work),
consistent with no variation.   The maser and 20 GHz continuum 
emission are coincident to within 40 pc (1$\sigma$; Table \ref{tab:positions} and Figure \ref{fig:SDSSJ0804}), 
which is less precise than the other objects due to the much larger distance and lower observing frequency.  The 12--16 GHz spectral 
index is $-1.5\pm0.3$ (Table \ref{tab:continuum}).  This object remains an inclined maser disk candidate.

\subsubsection{CGCG 120$-$039}
The water maser in this little-studied galaxy was detected in 2013 in GBT program
 GBT13A-236.\textsuperscript{\ref{footnote:maser_url}}
It had a peak flux density of $\sim$210 mJy, was blueshifted from the systemic velocity, and 
showed several components.  
We detect the maser --- now at $82\pm2$ mJy peak, but still showing multiple components (Figure \ref{fig:CGCG120-039})
--- and the 20 GHz continuum at $0.56\pm0.04$ mJy, which are spatially coincident to within a remarkably small $3.5\pm1.4$ pc 
(Table \ref{tab:positions}).  
The 18--22 GHz spectral index is flat:  $\alpha = -0.27\pm0.16$ (Table \ref{tab:continuum}).
This object is a good inclined maser disk candidate.  

\subsubsection{J0912+2304}  
The water maser in the galaxy J0912+2304 was detected in 2008 by GBT program GBT07A-034,\textsuperscript{\ref{footnote:maser_url}}
showing a $\sim$30 mJy peak.  \citet{zhu2011} list this as a water maser nondetection.  The VLA observations show a $16\pm3$ mJy 
peak, but no 20 GHz continuum down to an rms noise of 16 $\mu$Jy beam$^{-1}$ (Figure \ref{fig:2MASX0912} and Tables \ref{tab:obs} and
\ref{tab:masers}).  The provenance of the maser remains ambiguous.

\subsubsection{J1011$-$1926}  
The maser in this almost unknown galaxy was detected in GBT program GBT07A-066\textsuperscript{\ref{footnote:maser_url}}
in 2008.  \citep{zhu2011} list this object as a water maser nondetection.  
The GBT detection shows a broad line with a $\sim$80 mJy peak.  The VLA detection shows a broad multi-component maser with a 
$44\pm4$ mJy peak in good agreement with the systemic velocity (Figure \ref{fig:2MASX1011} and Table \ref{tab:masers}).  
We do not detect the 20 GHz continuum.  The nature of this maser therefore remains ambiguous.

\subsubsection{UGC 7016}   
The water maser in this barred spiral \citep{RC3} was 
detected in 2007 by GBT program GBT07A-034,\textsuperscript{\ref{footnote:maser_url}} but \citet{zhu2011} list this as a nondetection.
The GBT detection spectrum had a peak flux density of $\sim$55 mJy, but we did not detect the maser or the 20 GHz continuum emission.
VLA rms noise values were 3.8 mJy beam$^{-1}$ per 1.2 km s$^{-1}$ channel in the spectral line cube and 
16 $\mu$Jy beam$^{-1}$ in the continuum map (Table \ref{tab:obs}).

\subsubsection{NGC 5256}    
\citet{braatz2004} discovered the water maser in this merging luminous infrared galaxy in 2003.  
It had a peak flux density of 99 mJy ($L_{iso} = 30\ L_\odot$)  and was redshifted from the systemic
velocity by $\sim$300 km s$^{-1}$.  \citet{braatz2004} claim that all maser emission originates 
from the southern nucleus \citep[a Sey 2 nucleus according to, e.g.,][]{mazzarella1993}.
We detect two 20 GHz continuum sources that are consistent with the positions of the two nuclei 
\citep[Tables \ref{tab:positions} and \ref{tab:continuum};][]{brown2014},
but neither location (nor the larger region that includes the overlap region between the two nuclei)
shows maser emission down to an rms noise of 3.0 mJy beam$^{-1}$ in 1.2 km s$^{-1}$ channels
(Table \ref{tab:obs} and Figure \ref{fig:NGC5256}).
The spectral index of the northern nuclear continuum is $-2.0\pm0.2$ in the 18--22 GHz band  (Table \ref{tab:continuum}).

\subsubsection{NGC 5691}   
The water maser in NGC 5691, a barred non-Seyfert spiral galaxy \citep{mulchaey1997}, was detected in 
2009 by GBT program AGBT08C-035\textsuperscript{\ref{footnote:maser_url}}
with a $\sim$45 mJy peak flux density.  It was listed by \citet{zhu2011} as a maser nondetection.  
We did not detect the maser or any 20 GHz continuum in this galaxy, with rms noise levels of 3.7 mJy beam$^{-1}$ per 1.1 km s$^{-1}$ channel
and 17 $\mu$Jy beam$^{-1}$, respectively (Table \ref{tab:obs}).

\subsubsection{CGCG 168$-$018}  \label{subsubsec:CGCG168-018}
CGCG 168$-$018 is a little-studied galaxy classified as an AGN by \citet{schawinski2010} and
listed as a water maser nondetection by \citet{zhu2011}.
Water maser emission was detected by GBT program GBT07A-066\textsuperscript{\ref{footnote:maser_url}}
in 2008.  The GBT water maser spectrum shows 
a $\sim$50 mJy peak, detected by this work at $25\pm2$ mJy (Table \ref{tab:masers}).  
The $0.31\pm0.03$ mJy continuum shows a 18-22 GHz spectral index of $-0.95\pm0.17$ (Table \ref{tab:continuum}).   
The maser emission and 20 GHz continuum
are unresolved but show a significant relative offset of $21.6\pm2.7$ pc (Table \ref{tab:positions} and Figure \ref{fig:CGCG168-018}).  
This is the only object in the sample
that shows a significant offset between the maser and continuum emission (Figure \ref{fig:offsets}).
While the offset suggests that the maser emission is not deflected from an inclined maser disk, the continuum 
spectral index suggests that the continuum may arise from a jet, and the radio core is not detected.  The provenance 
of the maser therefore remains ambiguous.

\subsubsection{J1939$-$0124}   
\citet{greenhill2003} detected the water maser in this spiral galaxy hosting a Sey 2 nucleus in 2002 using the Tidbinbilla antenna.  
The maser showed a peak flux density of $\sim$28 mJy, and \citet{henkel2005} report $L_{iso} \simeq 160\ L_\odot$. 
No 22 GHz continuum was detected by \citet{greenhill2003} using the VLA ($< 2.8$ mJy).  
We likewise detect no continuum, with rms noise of 16 $\mu$Jy beam$^{-1}$ (Table \ref{tab:obs}), but we do detect the 
maser emission with peak $30\pm2$ mJy ($L_{iso} = 102\pm16 \ L_\odot$) although with a substantially different maser profile 
(Figure \ref{fig:2MASX1939} and Table \ref{tab:masers}).  
The 6 and 20 cm continua were detected at $5.4\pm0.4$ mJy  and $15.5\pm1.0$ mJy, respectively, by \citet{vader1993}, and the 
6 cm continuum position agrees with the maser position to within $\sim$1\arcsec ($\sim$430 pc).  
The nature of this maser remains ambiguous.

\section{Conclusions}

This paper presents a physical mechanism that may enable detection of inclined water maser disks orbiting massive black holes
via the lensing/deflection of in-going systemic masers.  The observational signature of an inclined disk 
is a maser line or line complex with limited Doppler extent that appears to arise at the location of the black 
hole, as identified by its radio continuum core.  With enough angular resolution, it may be possible to 
measure the black hole mass if the maser emission forms a lensing arc or Einstein ring, but the mass
precision will be limited by one's ability to measure or estimate the size of the maser-emitting portion of the disk.  

We suggest that if inclined maser disks can be detected at all, then they have probably already been detected
in single-dish surveys but discarded for interferometric follow-up because they did not show high-velocity lines.
We present original 0.07--0.17\arcsec\ (4--100 pc) resolution VLA observations of inclined maser disk candidates with the goal
of identifying systems where the maser emission is unresolved and is coincident with the 20 GHz continuum 
emission.  

Of the 16 masers observed with the VLA, we obtained useful data for 13, and among these, five were 
detected in both 22 GHz maser line emission and in 20 GHz continuum.  Of these five, one maser is most likely
associated with star formation (NGC 520b), and one shows a significant spatial offset between the maser emission 
and the continuum (CGCG 168$-$018, but it could still host an inclined maser disk --- this case is ambiguous).  
Three objects are good inclined maser disk candidates that merit further study with VLBI:  
IC 485, J0804+3607, and CGCG 120$-$039.
Five maser hosts remain ambiguous, based either on non-detected or offset continua:
J0350$-$0127,  J0912+2304, J1011$-$1926, CGCG 168$-$018, and J1939$-$0124.  

More straightforward methods for measuring black hole masses from molecular lines may be in the offing.  For example,
\citet{davis2013} and \citet{barth2016a,barth2016b} have used carbon monoxide kinematics in thin disks that approach or are
within the black hole gravitational sphere of influence to obtain black hole mass measurements.  \citet{barth2016a}, in particular, 
demonstrate the ability of ALMA to measure black hole masses with $\sim$10\% uncertainty.   Although they lack the 
intrinsic brightness of masers that enables VLBI mapping, 
thermal molecular lines have the advantage of being observable at any disk inclination.

\acknowledgments

We thank A. Hamilton and M. Eracleous for helpful discussion and B. Butler for assistance with 
metadata repair of bespoke observing configurations.  
We also thank the anonymous referee for important feedback.
This research has made use of NASA's Astrophysics Data System Bibliographic Services and
the NASA/IPAC Extragalactic Database (NED),
which is operated by the Jet Propulsion Laboratory, California Institute of Technology,
under contract with the National Aeronautics and Space Administration.

 \facility{VLA}

\software{CASA}

\end{document}